\documentclass[twocolumn,trackchanges]{aastex631}
\usepackage{amsmath}
\usepackage{enumitem}
\usepackage{wrapfig}
\usepackage[utf8]{inputenc}
\usepackage{array,multirow,graphicx}

\received{-}
\revised{-}
\accepted{-}

\graphicspath{{./}}

\begin{document}

\title{Large-scale cosmic ray anisotropies with 19 years of data from the Pierre Auger Observatory}
\collaboration{0}{The Pierre Auger Collaboration}
\email{spokespersons@auger.org}

\begin{abstract}
Results are presented for the measurement of large-scale anisotropies in the arrival directions of ultra-high-energy cosmic rays detected at the Pierre Auger Observatory during 19 years of operation, prior to AugerPrime, the upgrade of the Observatory. The 3D dipole amplitude and direction are reconstructed above $4\,$EeV in four energy bins. Besides the established dipolar anisotropy in right ascension above $8\,$EeV, the Fourier amplitude of the $8$ to $16\,$EeV energy bin is now also above the $5\sigma$ discovery level. No time variation of the dipole moment above $8\,$EeV is found, setting an upper limit to the rate of change of such variations of $0.3\%$ per year at the 95\% confidence level. Additionally, the results for the angular power spectrum are shown, demonstrating no other statistically significant multipoles. The results for the equatorial dipole component down to $0.03\,$EeV are presented, using for the first time a data set obtained with a trigger that has been optimized for lower energies. Finally, model predictions are discussed and compared with observations, based on two source emission scenarios obtained in the combined fit of spectrum and composition above $0.6\,$EeV.
\end{abstract}


\section{Introduction}

In the last decade, significant progress has been made in the search for the origin of ultra-high-energy cosmic rays (UHECRs) thanks to the data from the Pierre Auger Observatory \citep{AugerNIM2015} and Telescope Array \citep{TelescopeArray:2012uws}. The study of the arrival directions is a key element in gaining insight into the possible sources of UHECRs. At small and intermediate angular scales, interesting hints of anisotropies have been reported, the most significant being the overdensity in the Centaurus region, with a significance of $4\sigma$  (see \citep{PierreAuger:2022axr,PierreAuger:2023fcr} for the latest results). Regarding large angular scales, a dipolar modulation in right ascension (R.A.) at energies above $8\,$EeV has been determined with a significance above $5\sigma$ \citep{Science2017}. The direction of this dipole, lying ${\sim}115^\circ$ away from the Galactic Center, suggests an extragalactic origin for the cosmic rays above this energy threshold\footnote{Due to the smaller statistics of Telescope Array ($\sim 10$ times smaller), a large-scale dipolar anisotropy has not been confirmed in their data set, with a $99\%$ C.L. upper limit on the first-harmonic amplitude of $7.3\%$ 
\citep{TelescopeArray:2020cbq}.}.

Furthermore, an approximately linear growth of the dipole amplitude as a function of energy has been measured \citep{2018ApJ...868....4A}. This could be a consequence of the energy losses that cosmic rays suffer in interactions with the background radiation, that lead to a decrease of the horizon of cosmic rays sources  as the energy increases \citep{1966PhRvL..16..748G, 1966JETPL...4...78Z}. This shrinking of the horizon is expected to induce an increase in the dipole amplitude as a function of energy, due to the growing relative contribution of nearby sources, whose distribution is more inhomogeneous. In the alternative scenario in which only one or a few nearby sources give the dominant contribution to the flux above 4\,EeV, a rising dipolar amplitude with energy is also expected due to the growth of the UHECR magnetic rigidity, which implies smaller deflections of the particles in the intervening magnetic fields\footnote{For the source model that best fits the spectrum and composition results obtained in \citep{PierreAuger:2022atd}, the CR horizon is shrunk from a few hundreds of Mpc to a few tens of Mpc between $10\,$EeV and $100\,$EeV, and the mean rigidity changes from $\sim4\,$EV to $\sim 8\,$EV in the same energy range.}.

For energies below 4\,EeV, where only the anisotropies in right ascension can be studied, small amplitudes of the equatorial dipole, below the $1\%$ level, have been measured, which are compatible with isotropic expectations for the present accumulated statistics \citep{2020ApJ...891..142A}. However, a change in the right ascension phase at energies below 4~EeV  is apparent, with the observed values being close to the right ascension of the Galactic Center. This change could be indicative of a transition in the origin of the anisotropies from an extragalactic one at energies above a few EeV to a Galactic one at lower energies, or alternatively to the effects of the Galactic magnetic field on an extragalactic flux \citep{Mollerach:2022aji}. 

In this work, to gain further insight into these results, the statistics of 19 years of data from the Pierre Auger Observatory are analyzed, which represents an increase of ${\sim} 55\%$ with respect to \citep{2018ApJ...868....4A} and of ${\sim} 30\%$ with respect to \citep{2020ApJ...891..142A}.

An update on the angular power spectrum \citep{PierreAuger:2016gkp} is also presented to understand whether higher multipoles are present. Note that even a pure dipolar distribution at the boundaries of the Galaxy could be transformed, due to the deflections of cosmic rays in the Galactic magnetic field, into a dipolar distribution plus higher-order multipoles.

Finally, predictions are included, based on our combined fit of spectrum and composition data \citep{PierreAuger:2022atd}, for the dipolar amplitude and direction as well as for the quadrupolar amplitude. 

\section{The Pierre Auger Observatory and the data sets}

The Pierre Auger Observatory \citep{AugerNIM2015} is the largest cosmic ray observatory in the world, covering an area of $3000\,$km$^2$. It is located near the city of Malarg\"ue, Argentina (35.2$^\circ$~S, 69.5$^\circ$~W, 1400\,m~a.s.l.). It is a hybrid detector, consisting of a surface detector (SD) with 1660 water-Cherenkov stations and a fluorescence detector (FD) with 27 telescopes overlooking the array. The duty cycle of the SD is ${\sim}100\%$, while that corresponding to the FD, which operates on clear moonless nights, is ${\sim}15\%$. The main advantage of having both detectors is that the quasi-calorimetric energy determination obtained with the FD can be used to calibrate the energy estimate of the events registered with the SD, using the hybrid events  detected simultaneously by the SD and the FD. 

The main surface array (SD-1500) consists of 1600 detectors on a triangular grid with a spacing of $1500\,$m. There is a denser ``infilled array'' (SD-750) of 60 detectors with a spacing of $750\,$m, used to register events down to an energy threshold of ${\sim}0.03\,$EeV. There is also a smaller array with a spacing of $433\,$m, but the dataset has smaller statistics than the SD-750 and it is not included in this work. The events recorded with the SD-1500 that are considered here were registered between 1st January 2004 and 31st December 2022. For the transitional years of 2021 and 2022, when the AugerPrime \citep{Berat:2023ttm} installation was underway, only those detectors in which the electronics had not been updated are used (resulting in an equivalent $\sim$1.6 years of exposure instead of 2 years). 

The analyses above $4\,$EeV are made considering events with zenith angle $\theta< 80^\circ$, achieving an $85\%$ coverage of the sky. The events are selected if the SD station with the largest signal is surrounded by at least five active stations and if the reconstructed core of the shower falls within an isosceles triangle of nearby active stations. To  accurately compute the exposure as a function of time, those events that were registered during periods of unreliable data acquisition are removed, resulting in a total exposure of $123{,}000\,$km$^2\,$sr$\,$yr. There are actually different  reconstructions  for the events with $\theta< 60^\circ$ (``vertical'') and $60^\circ<\theta< 80^\circ$ (``inclined''). Full efficiency is reached above $2.5\,$EeV for the former events and above $4\,$EeV for the latter ones.

For studies using the SD-1500 array between $0.25\,$EeV and $4\,$EeV only vertical events are considered (leading to a $71\%$ coverage of the sky) and a stricter quality cut, requiring that the SD station with the largest signal should be surrounded by six active stations, is used. This dataset has an exposure of $81{,}000\,$km$^2\,$sr$\,$yr. 

Events with energies down to $0.03\,$EeV with the SD-750 array are also used. In this dataset, not only events registered using the standard station-level trigger algorithms are considered but also, for the first time, those detected with two other triggers introduced in mid-2013 to enhance the sensitivity of the array to small signals \citep{PierreAuger:2021hun}. Thanks to these triggers, full efficiency for the SD-750 array is reached above $0.2\,$EeV for events with $\theta<55^\circ$. The accumulated exposure of the SD-750 array with $\theta<55^\circ$ between 1st January 2014 and 13th December 2021 is $269\,$km$^2\,$sr$\,$yr.

The statistical uncertainty of the energy is ${\sim} 7\%$ for $E>10\,$EeV and can be up to ${\sim} 20\%$ for $E{\sim}0.1\,$EeV, while the systematic uncertainty of the absolute energy scale is $14\%$ \citep{2020PhRvD.102f2005A, PierreAuger:2021hun}. The events have an angular resolution better than $0.9^\circ$ for $E>10\,$EeV, which can degrade to $\lessapprox 2^\circ$ at lower energies \citep{AugerNIM2015,PierreAuger:2023onx}.

The energies of vertical events are corrected for atmospheric \citep{PierreAuger:2017vtr} and geomagnetic effects \citep{PierreAuger:2011yxe}, so as to avoid spurious modulations in the distributions in R.A. and azimuth, respectively. For inclined events, the air shower cascades near ground level are composed mostly of muons, and the atmospheric effects are negligible for them, while the geomagnetic effects are directly included in the reconstruction.

\section{Description of the large-angular-scale analyses}

In this section, the methods used in the present work are briefly discussed. For further details see \citep{PierreAuger:2014ati, 2018ApJ...868....4A,  PierreAuger:2016gkp, 2020ApJ...891..142A}.

\subsection{3D dipole above 4 EeV}

To reconstruct the 3D dipole, a separate Fourier analyses in R.A. ($\alpha$) and azimuth ($\phi$) is performed. The amplitude $r_k^x$ and phase $\varphi_k^x$ of the event rate modulation are given by
\begin{eqnarray}
r_k^x &= \sqrt{(a_k^x)^2+(b_k^x)^2}, \ 
\varphi_k^x&=\frac{1}{k} \arctan \frac{b_k^x}{a_k^x},
\end{eqnarray}
where the harmonic amplitudes of order $k$ are 
\begin{eqnarray}
a_k^x &= \frac{2}{\mathcal{N}} \sum_{i=1}^N w_i \cos(kx_i), \ 
b_k^x &= \frac{2}{\mathcal{N}} \sum_{i=1}^N w_i \sin(kx_i),
\end{eqnarray}
with $x=\alpha$ or $\phi$ and $k=1$ for the dipolar component. $N$ is the total number of events, $w_i$ are the weight factors and $\mathcal{N}=\sum_{i=1}^N w_i$ is the normalization factor. The weight factors $w_i$ are computed as
\begin{equation}
w_i=\left [\Delta N_{{\rm cell}}(\alpha^0(t_i))(1+0.003 \tan \theta_i \cos(\phi_i-\phi_0)) \right ]^{-1},
\end{equation}
where $\Delta N_{{\rm cell}}(\alpha^0(t_i))$ is the relative variation of the total number of active detector cells for a given R.A. of the zenith of the Observatory $\alpha^0$ evaluated at the time $t_i$ at which the $i$th event is detected, and $\theta_i$ and $\phi_i$ are the zenith and azimuth of the event, respectively. These weights are used to take into account the slightly non-uniform exposure of the array over time (due to the deployment of the array and the sporadic downtime of the detectors) and the tilt of the array (which is on average tilted $0.2^\circ$ towards $\phi_0=-30^\circ$, with $\phi_0$ measured anticlockwise from the East). Both effects could induce spurious modulations if they are not accounted for. 

The probability that an amplitude equal or larger than $r_k^\alpha$ arises from a fluctuation from an isotropic distribution of events is given by $P(\ge r_k^\alpha)= \exp(-\mathcal{N}(r_k^\alpha)^2/4)$ \citep{Linsley:1975kp}.

Assuming that the dominant anisotropy in the distribution of arrival directions of cosmic rays is given by the dipolar component, $\mathbf{d}$, the flux distribution can then be written as  
\begin{eqnarray}
 \Phi(\hat{u}) &=&\Phi_0 \left ( 1 + \Delta(\hat{u}) \right ), \nonumber \\
\Delta(\hat{u}) &=& \mathbf{d} \cdot \hat{u},
\end{eqnarray}
where $\hat{u}$ is the arrival direction of the cosmic rays. The equatorial amplitude of the dipole, $d_\perp$, the North-South component, $d_z$, and the dipole direction in Equatorial coordinates ($\alpha_d$,\,$\delta_d$) are related to the first-harmonic amplitudes in R.A. and azimuth through
\begin{eqnarray}
d_\perp &\simeq \frac{r_1^\alpha}{\left<\cos \delta \right>}, \ 
d_z &\simeq \frac{b_1^\phi}{\cos l_{\rm obs} \left<\sin \theta \right>}, \nonumber\\
\alpha_d &= \varphi_1^\alpha, \ 
\delta_d &= \arctan \left( \frac{d_z}{d_\perp}\right), \label{3ddipoleFourier}
\end{eqnarray}
where $\left<\cos \delta \right>  \approx 0.7814$ is the mean cosine of the declination of the events, $\left<\sin \theta \right>  \approx 0.6525$ is the mean sine of the zenith of the events and $l_{\rm obs} = -35.2^\circ$ is the latitude of the Observatory.

\subsection{3D dipole and quadrupole above 4 EeV}

In case that a quadrupolar distribution is also present, the flux anisotropy can be parameterized as 
\begin{equation}
 \Delta(\hat{u})=\ \mathbf{d} \cdot \hat{u} + \frac{1}{2} \sum_{i,j} Q_{ij} u_i u_j
\end{equation}
where $Q_{ij}$ are the components of the quadrupole tensor (five of them are independent). The details on how the dipole and quadrupole components are estimated from the $k =1$ and $2$ harmonic amplitudes are described in \citep{PierreAuger:2014ati}.

\subsection{Angular power spectrum above 4 EeV}

To search for anisotropies across various angular scales, it is convenient to decompose the distribution of observed events per unit solid angle $dN/d\Omega(\hat{u})$ in each direction $\hat{u}$, separating the dominant isotropic contribution from the anisotropic component, $\Delta(\hat{u})$, as
\begin{equation}\label{eq:fluxo_parcial}
    \frac{dN}{d\Omega}(\hat{u}) = \frac{N}{4\pi f_1} {\rm{W}}(\hat{u})\left ( 1 + \Delta(\hat{u}) \right ), 
\end{equation}
where ${\rm W}(\hat{u})$ is the relative exposure of the observatory, $N$ is the total number of observed events and  $f_1 =\frac{1}{4\pi} \int {\rm W}(\hat{u}) {\rm d}\hat{u}$. 
The angular power spectrum of the scalar field ${\rm\Delta}(\hat{u})$ is defined by $C_{\ell}= \frac{1}{2\ell + 1}\sum_{m=-\ell}^{\ell} |a_{\ell m}|^2$, where the coefficients $a_{\ell m}$, encoding any anisotropy signature present in data, are derived from the multipolar expansion of $\Delta(\hat{u})$ into spherical harmonics, $a_{\ell m} = \int \Delta(\hat{u}) Y_{\ell m}(\hat{u}) d\hat{u}$.  Due to the incomplete sky coverage of the Pierre Auger Observatory, the estimation of the individual $a_{\ell m}$ coefficients cannot be carried out with relevant resolution as soon as $\ell_{\rm max} > 2$ and  the same is true for the power spectrum (full-sky analyses are carried out by the Pierre Auger and Telescope Array Collaborations together \citep{TelescopeArray:2014ahm, PierreAuger:2023mvf}). However, it is possible to reconstruct the angular power spectrum within a statistical resolution independent of the bound $\ell_{\rm max}$ \citep{Deligny2004}, if the observed distribution of arrival directions represents a particular realization of an underlying stochastic process in which the anisotropies cancel in the ensemble average $( \langle\Delta(\hat{u})\rangle = 0)$ and the second-order moment ($\langle\Delta(\hat{u})\Delta(\hat{u}')\rangle$) depends only on the angular separation between $\hat{u}$ and $\hat{u}'$. For a detailed discussion of the implications and restrictions of this hypothesis, see 
comments in the References section 
\citep{PierreAuger:2016gkp}. Under this hypothesis, the ensemble-average  expectations of the power spectrum $\left< C_\ell \right>$ and the ‘pseudo’-power spectrum $\left< \tilde{C}_{\ell} \right>$ are related  through

 \begin{equation}
\label{eqn:Cl0}
\left< \tilde{C}_{\ell} \right> = \sum_{\ell'}M_{\ell\ell'} \left< C_{\ell'} \right> + \frac{4 \pi f_1^2}{N}, 
\end{equation}with the ‘pseudo’-power spectrum $\tilde{C}_{\ell}=\sum^{\ell}_{m=-\ell}|\tilde{a}_{\ell m}|^{2}/(2\ell +1)$ defined in terms of the ‘pseudo’-multipolar moments $\tilde{a}_{\ell m} = \int {\rm{W}}(\hat{u})\Delta(\hat{u}) Y_{\ell m}(\hat{u}) d\hat{u}$. The operator $M$, describing the mixing between the modes, is entirely determined by the knowledge of the relative coverage function \citep{Deligny2004} and the $\frac{4 \pi f_1^2}{N}$ term  corresponds to an irreducible noise induced by Poisson fluctuations. Thus, for a cosmic-ray dataset with a ‘pseudo-power’ spectrum $\tilde{C}_{\ell}$,  the angular power spectrum  $\hat{C}_\ell$\footnote{In \citep{PierreAuger:2016gkp}, unlike in this work, the results were presented without removing the noise term. Thus, by construction, in this work, the average of the isotropic simulations is zero, and any particular realization with a negative $C_{\ell}$ is interpreted as a statistical fluctuation consistent with isotropy.}, which is an unbiased estimator in the ensemble average, is computed by

 \begin{equation}
\label{eqn:Cl}
\hat{C}_{\ell} = \sum_{\ell'}M_{\ell\ell'}^{-1}  \tilde{C}_{\ell'} - \frac{4 \pi}{N} \frac{f_1^2}{f_2},
\end{equation}with $f_2 =\frac{1}{4\pi} \int {\rm W}^2(\hat{u}) {\rm d}\hat{u}$.

\subsection{Modulation in R.A. down to 0.03 EeV}

When the trigger efficiency of the array is small, the systematic effects on the azimuth distribution cannot be completely corrected for, in particular the dependence of the trigger efficiency on azimuth due to the geometry of the layout of the array. Thus, for lower energies we restrict the large-scale studies to the modulation in R.A., which is proportional to the equatorial dipole component as shown in Eq.~\ref{3ddipoleFourier}. 

Due to the Earth's rotation, the exposure of the Observatory is almost uniform in R.A. and above full-efficiency the sources of spurious modulation (sporadic detector downtime and atmospheric effects) can be corrected for as described above. The effect of the tilt of the array is not relevant for the Fourier analysis in R.A. (it only affects the distribution in azimuth), thus it is not included in the weights for these analyses.

Trigger effects at energies where the efficiency is low are difficult to control down to the required precision for reconstructing per mille level anisotropies. However, a method suitably constructed to be insensitive to such effects, called the East-West method, \citep{Bonino:2011nx} can be used. This method, which is based on the difference between the counting rates of the events detected from the east sector and those from the west sector, is less sensitive than the Fourier method, but the systematics are under better control. The Fourier coefficients for the East-West method are
\begin{eqnarray}
a_{\rm EW} &= \frac{2}{N} \sum_{i=1}^{N} \cos(\alpha^0(t_i)-\xi_i), \nonumber\\
b_{\rm EW} &= \frac{2}{N} \sum_{i=1}^{N} \sin(\alpha^0(t_i)-\xi_i),
\end{eqnarray}
where $\alpha^0(t_i)$ is the R.A. of the zenith of the array, and $\xi_i=0$ for events coming from the East ($-\pi/2<\phi<\pi/2$) and $\xi_i=\pi$ for those coming from the West ($\pi/2<\phi<3\pi/2$).

The Fourier amplitude, $r$, and phase, $\varphi$, are related to the East-West method values, $r_{\rm EW}$ and $\varphi_{\rm EW}$, through
\begin{eqnarray}
r &= \frac{\pi}{2} \frac{\left < \cos \delta \right >}{\left< \sin \theta \right>} r_{\rm EW} = \frac{\pi}{2} \frac{\left < \cos \delta \right >}{\left< \sin \theta \right>} \sqrt{a_{\rm EW}^2+b_{\rm EW}^2}, \nonumber\\
\varphi &= \varphi_{\rm EW} + \frac{\pi}{2} = \arctan \left (\frac{b_{\rm EW}}{a_{\rm EW}} \right) + \frac{\pi}{2}, 
\end{eqnarray}
as in \citep{Bonino:2011nx}.
As in the Fourier method, the probability of obtaining an amplitude larger than that expected from an isotropic distribution is $P(\ge r_{\rm EW})= \exp(-N(r_{\rm EW})^2/4)$. 

For the 0.25-$0.5\,$EeV energy bin, one can also use the data of the SD-750 array with the standard Fourier method since the SD-750 array is fully efficient at  these energies. The disadvantage of having lower statistics with the SD-750 array compared to those of the SD-1500 array at these energies is compensated by the fact that the number of events needed to have an equal statistical uncertainty with the Fourier method than with the East-West one is approximately a factor four smaller (the statistical uncertainty obtained with the East–West method is larger by a factor $\pi \left < \cos\delta \right >/2\left <\sin\theta \right > \sim 2.1$ for $\theta <60^\circ$ \citep{Bonino:2011nx}).

For the energy bins below $0.25\,$EeV, the statistics of the SD-750 array is larger than that of the main array, and the East-West method is applied given that for those energies the trigger of the SD-750 array is not fully efficient. For the analyses in R.A., the geomagnetic corrections are not necessary (they are only relevant for the analyses in azimuth) and, therefore, these corrections are not done for the SD-750 dataset. However, the atmospheric corrections in the reconstructed energy as well as the not completely uniform exposure in R.A. of the array are accounted for in the analyses with the SD-750 dataset, like in the SD-1500 dataset.

\section{Results}
\subsection{3D dipole above 4 EeV}

The results for the 3D dipole reconstruction above full efficiency are listed in Table~\ref{tab:3Ddipole}. For $E \ge 8\,$EeV, the significance of the dipolar modulation in R.A. is now at $6.8\sigma$ and its significance in the 8-$16\,$EeV energy bin is $5.7\sigma$. The uncertainties reported correspond to the $68\%$ confidence level (CL) of the respective marginalized probability distribution function (p.d.f).

\begin{figure*}[!ht]
\gridline{
          \fig{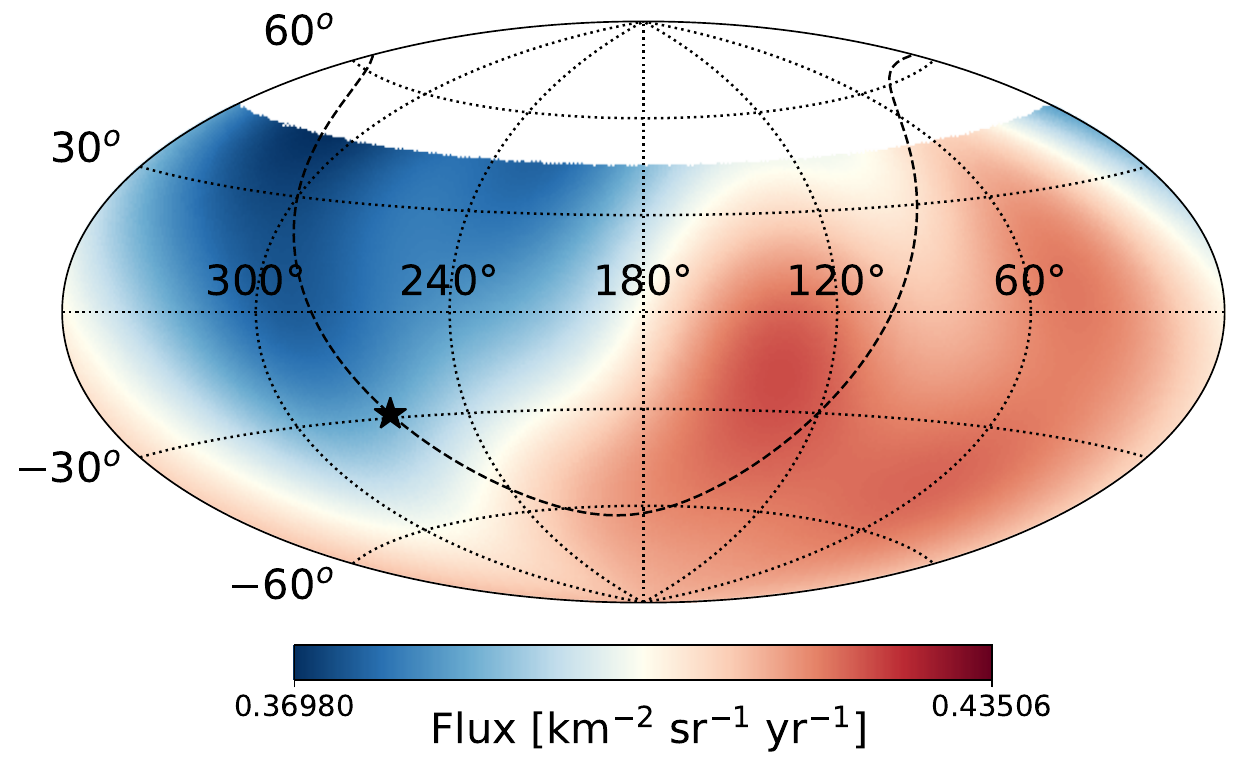}{0.45\textwidth}{(a)}
          \fig{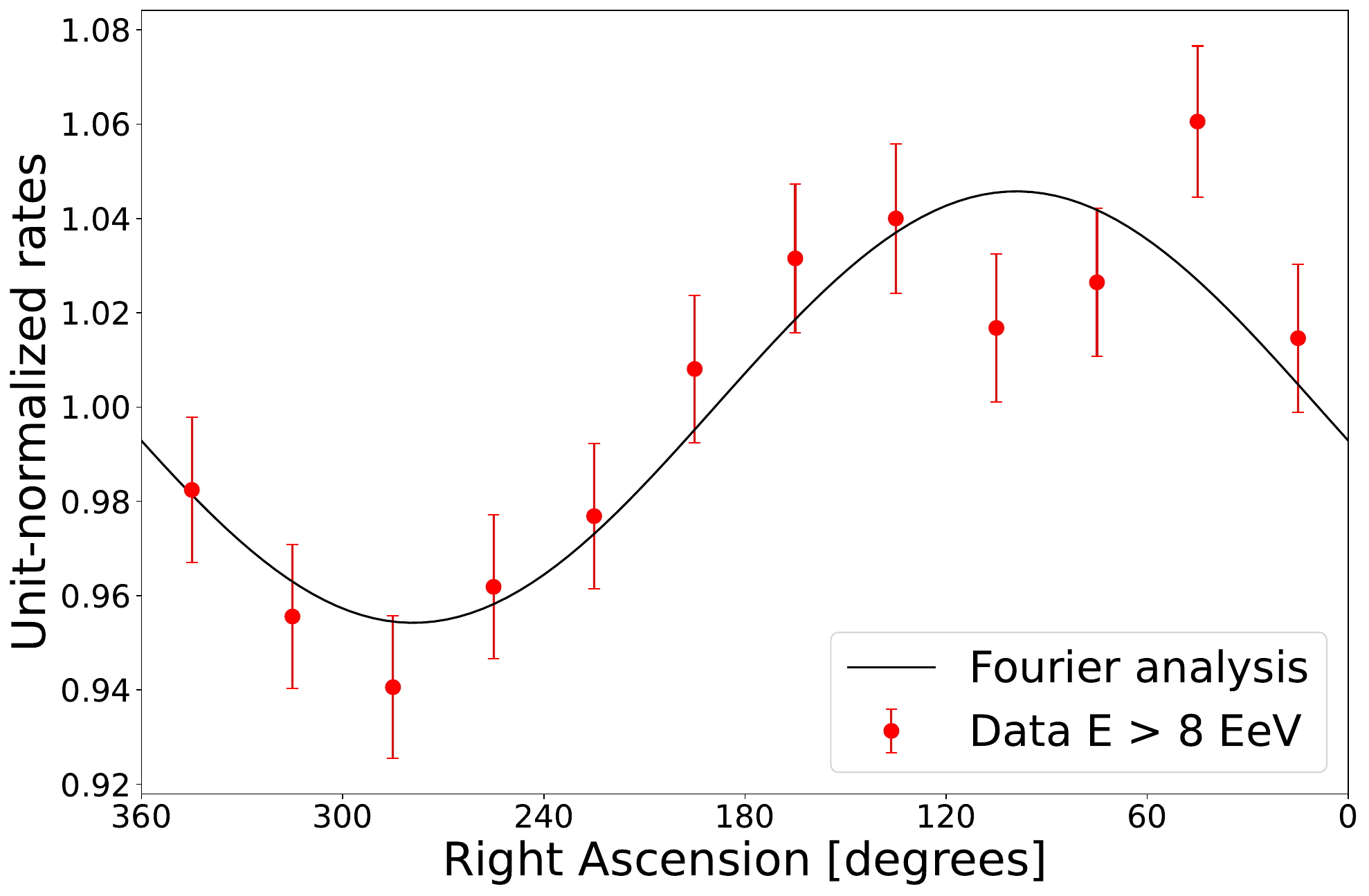}{0.4\textwidth}{(b)}
}
\caption{(a) Flux above $8\,$EeV, smoothed by a Fisher distribution with a mean cosine of the angular distance to the center of the window equal to that of a top-hat distribution with radius of $45^\circ$, in Equatorial coordinates. The position of the Galactic Center is shown with a star and the Galactic Plane is indicated with a dashed line. (b) Distribution in R.A. of the unit-normalized rates of events with $E \ge 8\,$EeV. The black line shows the obtained distribution with the Fourier analysis assuming only a dipolar component.}
    \label{fig:LSA_3Ddipole}
\end{figure*}

\begin{figure*}[!ht]
\gridline{
\fig{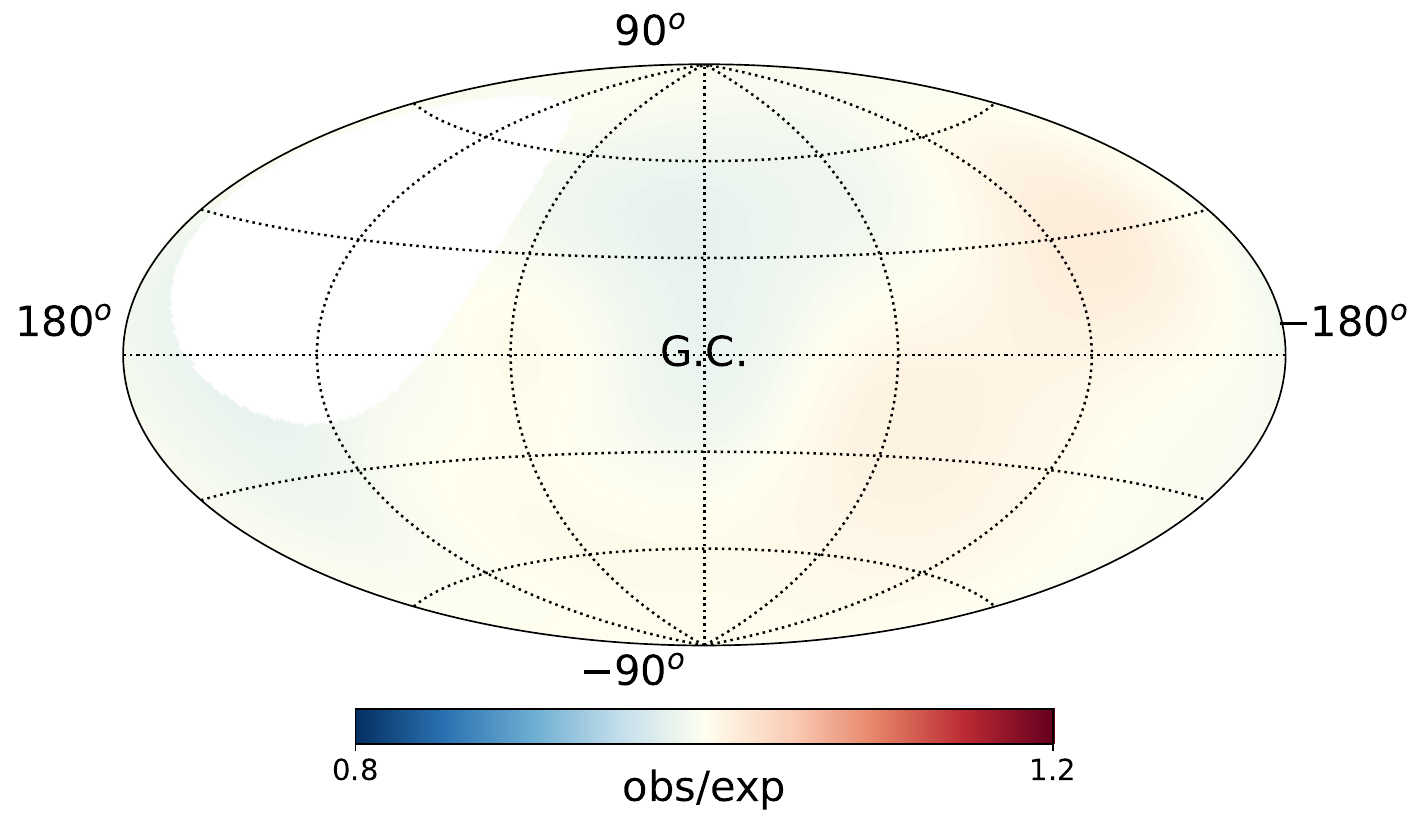}{0.48\textwidth}{(a)}
\fig{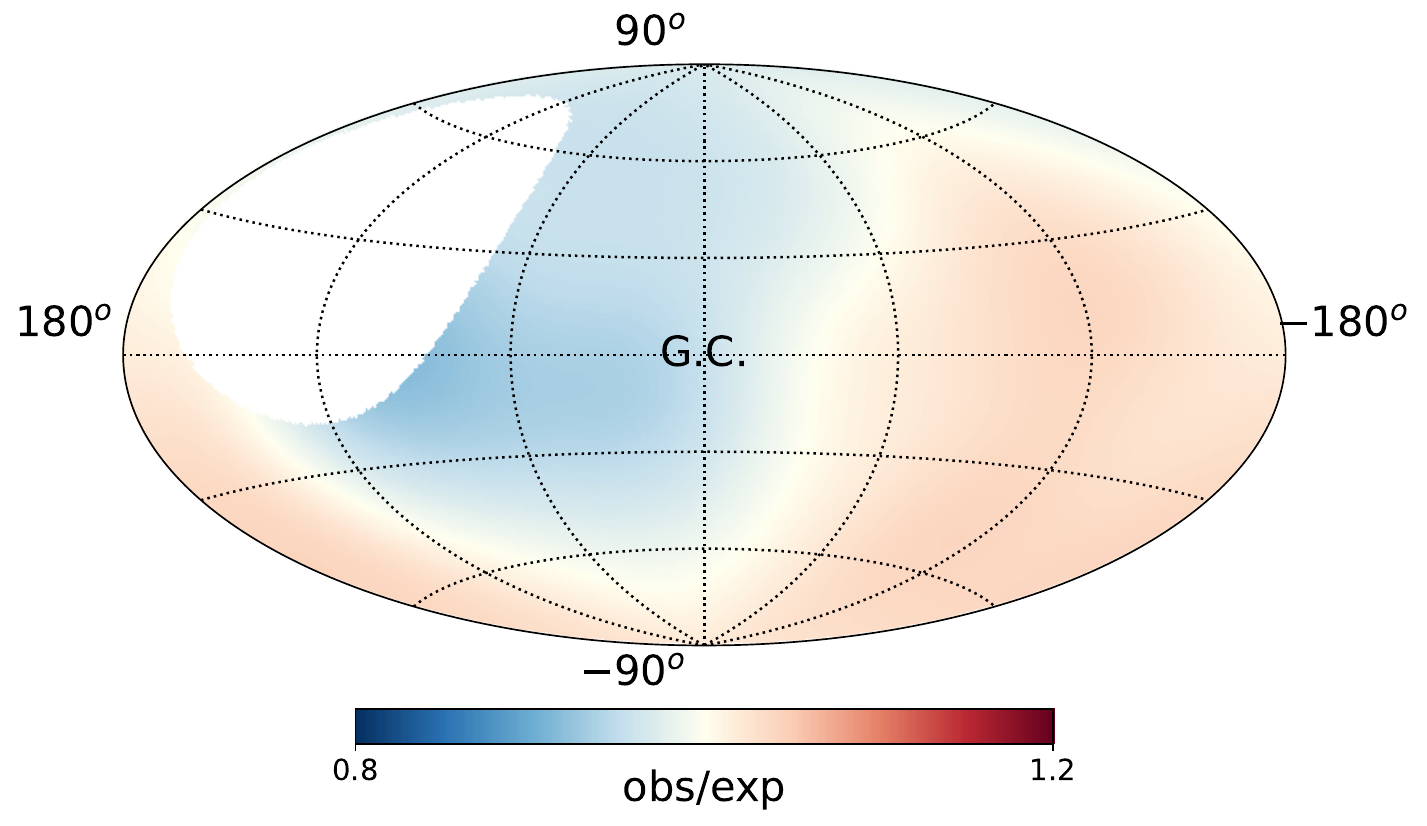}{0.48\textwidth}{(b)}}
\gridline{
\fig{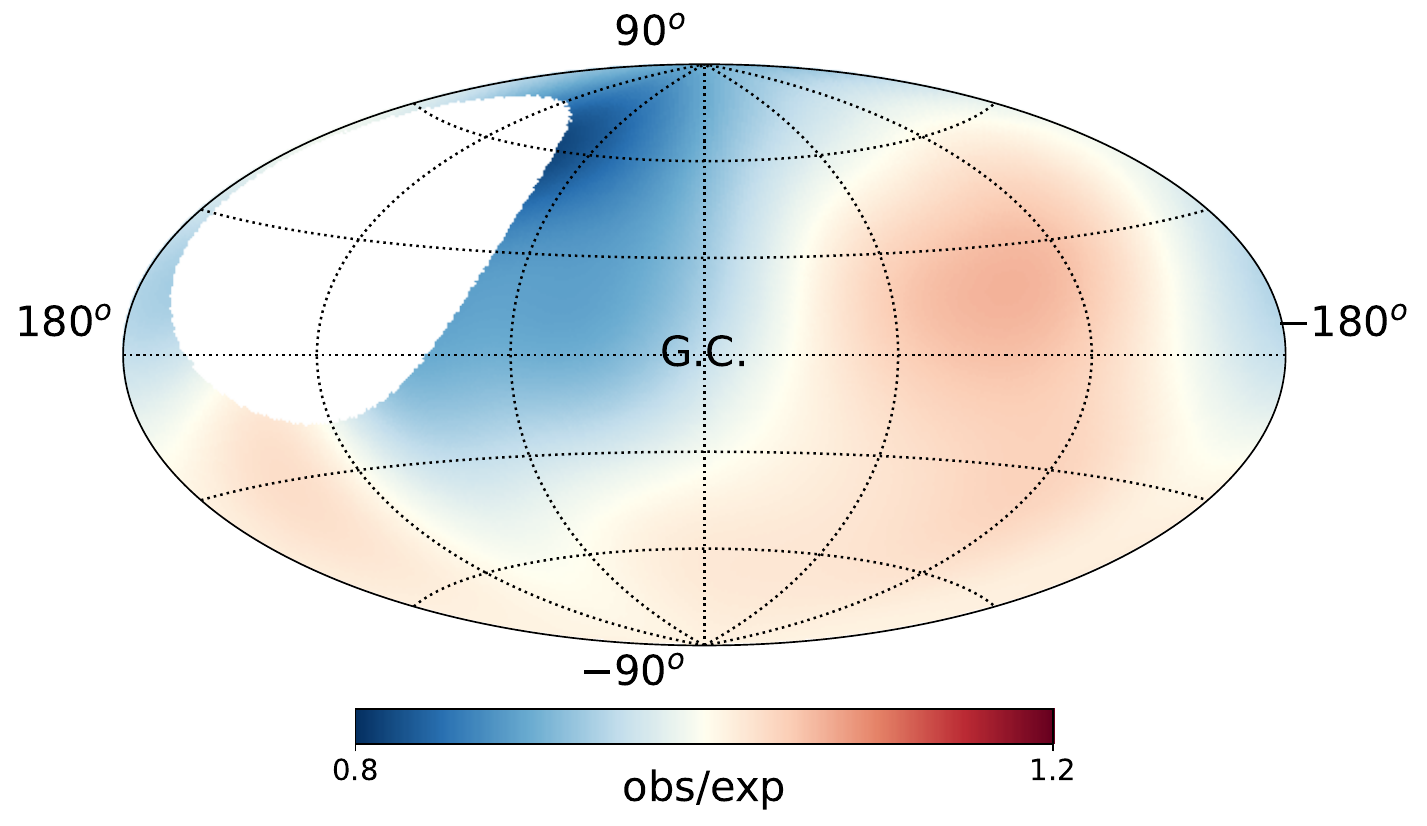}{0.48\textwidth}{(c)}
\fig{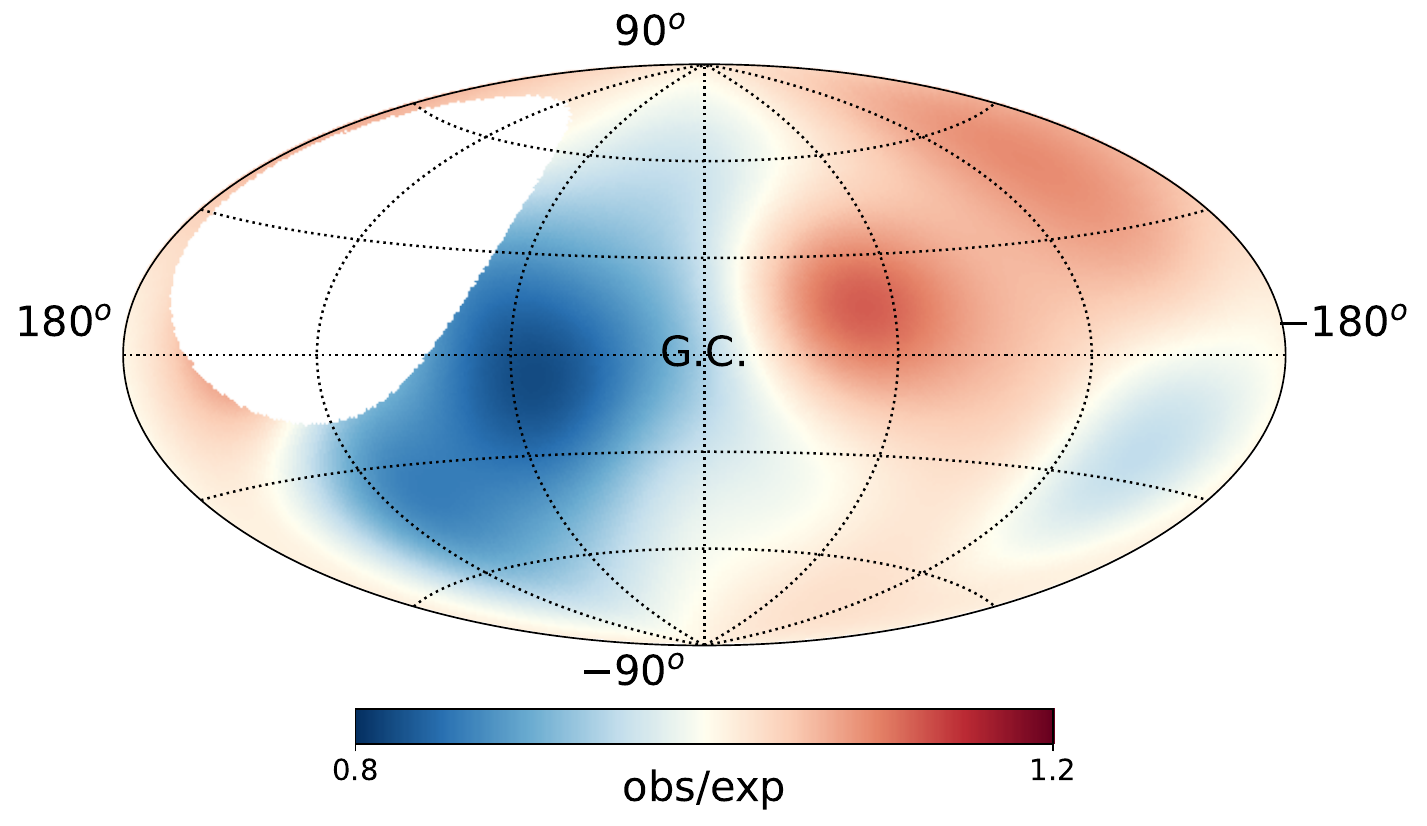}{0.48\textwidth}{(d)}
}
    \caption{Ratio between the number of observed events and those expected from an isotropic distribution, in Galactic coordinates, for the (4-8)\,EeV (a), (8-16)\,EeV (b), (16-32)\,EeV (c), and $\ge32$\,EeV (d) energy bins. The smoothing corresponds to a Fisher distribution with a mean cosine of the angular distance to the center of the window equal to that of a top-hat distribution with radius of $45^\circ$.}
    \label{fig:obs/exp_bins}
\end{figure*}

In Fig.~\ref{fig:LSA_3Ddipole}, the flux above $8\,$EeV in Equatorial coordinates and the distribution in R.A. of the rates of events (normalized to unity) for that energy threshold are depicted. In Fig.~\ref{fig:obs/exp_bins} sky maps in Galactic coordinates are included showing the ratio between the number of observed events and those expected from an isotropic distribution of arrival directions, for the energy bins of (4-8, 8-16, 16-32, $\ge32$)\,EeV. In Fig.~\ref{fig:obs/exp_bins} (d), the overdensity of events in the Centaurus region, ($l, b$)=($310^\circ,19^\circ$), is visible, contributing in the region corresponding to the dipole excess.

\begin{figure*}[!ht]
\gridline{\fig{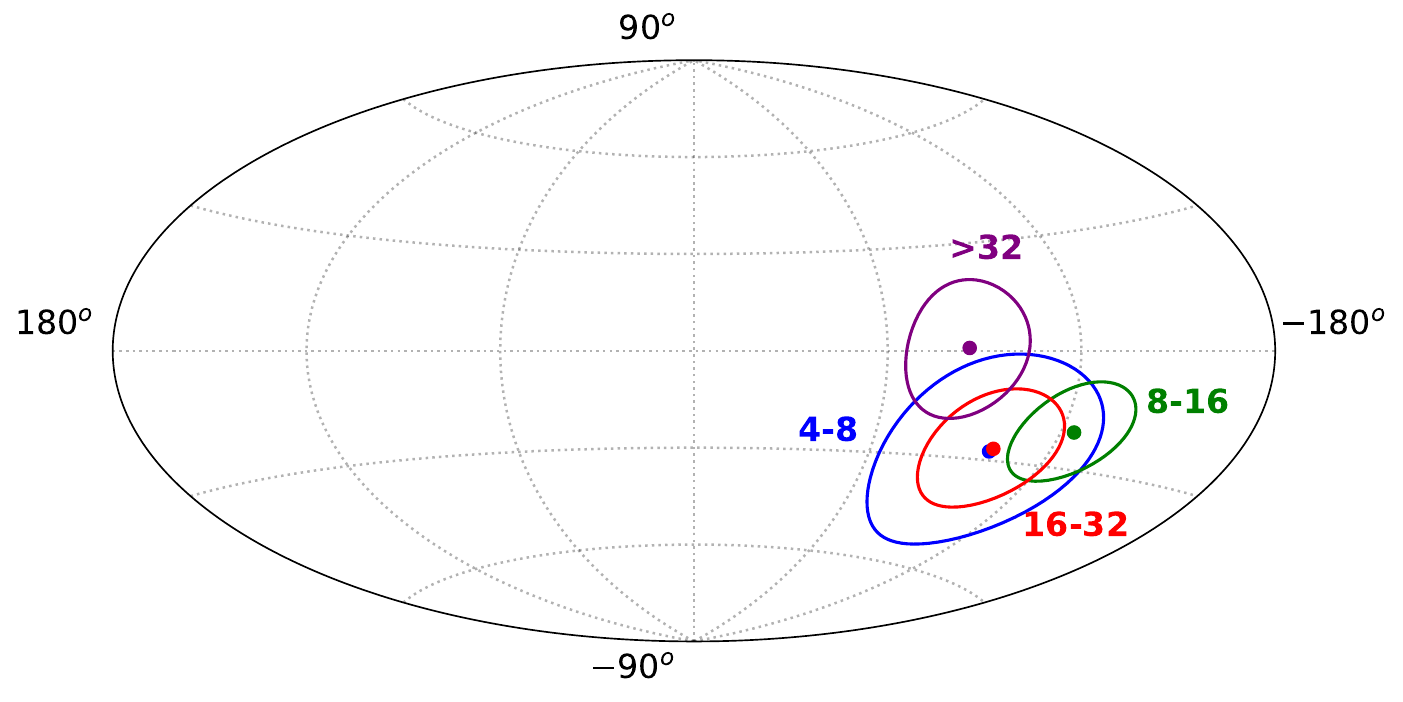}{0.6\textwidth}{(a)}
          \fig{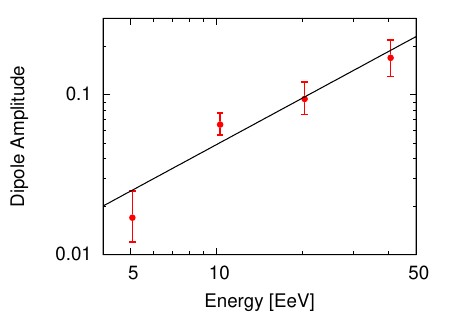}{0.4\textwidth}{(b)}
}
\caption{(a) Map with the directions of the 3D dipole for different energy bins, in Galactic coordinates. The contours of equal probability per unit solid angle, marginalized over the dipole amplitude, that contain the $68\%$ CL range are shown. (b) The evolution of the dipole amplitude with energy, for the four energy bins considered (4-8, 8-16, 16-32, $\ge32$)\,EeV.}
    \label{fig:LSA_dipoleEbin}
\end{figure*}

In Fig.~\ref{fig:LSA_dipoleEbin}, the evolution with energy of the dipole direction and amplitude is plotted. A fit to the amplitude as a function of energy, $d(E)=d_{10}\times \left( \frac{E}{10\ {\rm EeV}} \right)^\beta$, is performed obtaining $d_{10}=0.049 \pm 0.009$ and $\beta=0.97 \pm 0.21$, in agreement with \citep{2018ApJ...868....4A}. This growth in amplitude is possibly due to the larger relative contribution from the nearby sources for increasing energies, whose distribution is more inhomogeneous. Another effect which also results in an increase of the dipole amplitude with energy, secondary to this one, is the growth of mean rigidity of the particles, leading to smaller deflections and thus larger dipolar amplitudes (see Section 5 for a comparison of model predictions to data).

\begin{deluxetable*}{cccccccc}[!ht]
\tablecaption{Results for the 3D dipole reconstruction above full efficiency. For each energy bin the number of events, $N$, the equatorial component of the amplitude, $d_\perp$, the North-South component $d_z$, the amplitude $d$, the R.A., $\alpha_d$, and declination, $\delta_d$, of the dipole direction and the probability of getting a larger amplitude from fluctuations of an isotropic distribution $P(\ge r_1^\alpha)$ are presented. \label{tab:3Ddipole} }
\tablehead{
\colhead{$E$ [${\rm EeV}$]} & \colhead{$N$} & \colhead{$d_\perp$ [$\%$]} & \colhead{$d_z$ [$\%$]}	& \colhead{$d$ [$\%$]} &	\colhead{$\alpha_d$ [$^\circ$]} & \colhead{$\delta_d$ [$^\circ$]} & \colhead{$P(\ge r_1^\alpha)$}}
\startdata
    4-8 & 118,722 & $1.0^{+0.6}_{-0.4}$ & $-1.3\pm 0.8$ & $1.7^{+0.8}_{-0.5}$ & $92 \pm 28$ & $-52^{+21}_{-19}$ & 0.14 \\     
    $\ge$8 & 49,678 & $5.8^{+0.9}_{-0.8}$ & $-4.5\pm 1.2$ & $7.4^{+1.0}_{-0.8}$ & $97 \pm 8$ & $-38^{+9}_{-9}$ & $8.7 \times 10^{-12}$ \\         
    \hline         
    8-16 & 36,658 & $5.7^{+1.0}_{-0.9}$ & $-3.1\pm 1.4$ & $6.5^{+1.2}_{-0.9}$ & $93 \pm 9$ & $-29^{+11}_{-12}$ & $1.4 \times 10^{-8}$  \\
    16-32 & 10,282 & $5.9^{+2.0}_{-1.8}$ & $-7 \pm 3$ & $9.4^{+2.6}_{-1.9}$ & $93 \pm 16$ & $-51^{+13}_{-13} $ & $4.3 \times 10^{-3}$  \\    
    $\ge$32 & 2,738 & $11^{+4}_{-3}$ & $-13\pm 5$ & $17^{+5}_{-4}$ & $144 \pm 18$ & $-51^{+14}_{-14}$ & $9.8 \times 10^{-3}$  \\  
\enddata
\end{deluxetable*}

To show that there are no significant unaccounted systematics effects, in Table~\ref{table:solar_antis} the solar and anti-sidereal first-harmonic amplitudes ($r_1^{\rm solar}$ and $r_1^{\rm antis}$) and the probability of getting a larger amplitude from fluctuations of an isotropic distribution ($P(\ge r_1^{\rm solar})$ and $P(\ge r_1^{\rm antis})$) are reported, for each energy bin in which the Fourier analysis in R.A. was done at the sidereal frequency. The results are compatible with no significant modulation being present at these frequencies. Furthermore, for the Fourier analysis in azimuth, the $a^\phi_1$ parameters, which measure the East-West modulation, were verified to be compatible with zero (since any flux modulation from the sky should be averaged due to the Earth's rotation). 

\begin{table}[!ht]
\centering
\caption{Solar and anti-sidereal first-harmonic amplitudes, $r_1^{\rm solar}$ and $r_1^{\rm antis}$, and the corresponding isotropic probabilities, $P(\ge r_1^{\rm solar})$ and $P(\ge r_1^{\rm antis})$, for each energy bin. \label{table:solar_antis}}
\begin{tabular}{ccccc}
    \hline \hline
$E$ [${\rm EeV}$] & $r_1^{\rm solar}$ [\%] & $P(\ge r_1^{\rm solar})$ & $r_1^{\rm antis}$ [\%] & $P(\ge r_1^{\rm antis})$\\
\hline
2-4     & $0.4^{+0.3}_{-0.2}$ & 0.18 & $0.3^{+0.3}_{-0.1}$ & 0.48   \\ 
4-8     & $0.7^{+0.5}_{-0.3}$ & 0.28 & $0.4^{+0.5}_{-0.2}$ & 0.65   \\ 
$\ge8$  & $0.3^{+0.9}_{-0.04}$ & 0.91 & $1.4^{+0.7}_{-0.5}$ & 0.10   \\    
8-16    & $0.6^{+0.9}_{-0.2}$ & 0.71 & $1.1^{+0.8}_{-0.5}$ & 0.36   \\
16-32   & $2.2^{+1.6}_{-0.9}$ & 0.29 & $2.7^{+1.6}_{-1.0}$ & 0.15   \\
$\ge32$ & $1.6^{+3.5}_{-0.4}$ & 0.83 & $1.0^{+4}_{-0.01}$ & 0.93   \\  
\hline
\end{tabular}
\end{table}

The statistics of the $E\ge 8\,$EeV cumulative energy bin is large enough that the dataset can be separated in time-ordered subsets and tested for their stability. The results, dividing the dataset into two (the date corresponding to half the dataset is 08/19/2014) and four subsets (the dates corresponding to the end of each quarter of the data set being 09/15/2010, 08/19/2014, 09/05/2018 and 12/31/2022), are listed in Table \ref{table:time_bins}. It is seen that the equatorial dipole amplitude and phase are consistent within the statistical uncertainties in the different time periods considered. Given that no time variation of the equatorial dipole modulation is found, an upper limit to its long-term rate of change of 0.003 per year at the 95\% CL can be set using a linear fit to the split data, which illustrates the long-term stability of the detector. Another check that was performed was to separate the dataset into two zenith angle ranges, larger or smaller than $60^\circ$, corresponding to the different reconstructions (vertical and inclined) and the results are compatible with each other.

\begin{table}[!ht]
\centering
\caption{Number of events ($N$), reconstructed equatorial dipole amplitude ($d_\perp$), phase ($\alpha_d$) and isotropic probability ($P(\ge r_1^\alpha)$) separating the dataset into time-ordered subsets for the $E\ge 8\,$EeV cumulative energy bin. The first two rows correspond to separating the dataset into two time-ordered bins, while the following rows correspond to separating the dataset into four time-ordered bins. \label{table:time_bins}}
\begin{tabular}{ccccc}
\hline \hline
${\rm Time\ bins}$ & $N$ & $d_\perp (\%)$ & $\alpha_d$ [$^\circ$] & $P(\ge r_1^\alpha)$\\
\hline
1/2 & 24,839 & $5.5^{+1.3}_{-1.0}$ & $100 \pm 12$ & $1.1 \times 10^{-5}$ \\ 
2/2 & 24,839 & $6.1^{+1.3}_{-1.1}$ & $94 \pm 11$ & $6.9 \times 10^{-7}$ \\
\hline   
1/4 & 12,419 & $6.3^{+1.8}_{-1.4}$ & $111 \pm 15$ & $5.6 \times 10^{-4}$ \\ 
2/4 & 12,420 & $4.9^{+1.9}_{-1.3}$ & $87 \pm 19$ & $1.1 \times 10^{-2}$ \\
3/4 & 12,419 & $6.8^{+1.8}_{-1.4}$ & $92 \pm 14$ & $1.5 \times 10^{-4}$ \\ 
4/4 & 12,420 & $5.4^{+1.9}_{-1.4}$ & $97 \pm 17$ & $3.8 \times 10^{-3}$ \\
\hline
\end{tabular}
\end{table}

\subsection{3D dipole and quadrupole above 4 EeV}

In Table~\ref{tab:dip+quad}, the results obtained if a quadrupolar component is included in the Fourier analysis for each energy bin are shown. The dipole components, $\mathbf{d}=(d_x,d_y,d_z)$, the five independent quadrupolar components, $Q_{ij}$, the quadrupole amplitude, $Q \equiv \sqrt{\sum_{ij} Q_{ij}^2/9}$, and the $99\%$ CL upper limits, $Q^{\rm UL}$, are listed. It is seen that the quadrupolar components are not significant and that the dipole components are consistent with the results we obtain assuming only a dipole.

\begin{deluxetable*}{cccccc}[!ht]
\tablecaption{{Results obtained considering both dipolar and quadrupolar components. For every energy bin, the dipole components, $\mathbf{d}=(d_x,d_y,d_z)$, the five independent quadrupolar components, $Q_{ij}$, the quadrupole amplitude, $Q$, and the $99\%$ CL upper limit, $Q^{\rm UL}$, are presented. The x-axis is chosen along the $\alpha=0$ direction. \label{tab:dip+quad}}}
\tablehead{\colhead{} & \colhead{$4 \textendash 8\,{\rm EeV}$} & \colhead{$\ge 8\,{\rm EeV}$} & \colhead{$8 \textendash 16\,{\rm EeV}$} & \colhead{$16 \textendash 32\,{\rm EeV}$} & \colhead{$\ge 32\,{\rm EeV}$}}
\startdata
\hline
 $d_x$          & $-0.003\pm 0.007$ & $-0.002\pm 0.011$ & $-0.002\pm 0.012$ & $0.029\pm 0.024$  & $-0.1\pm 0.5$ \\
 $d_y$          & $0.005\pm 0.007$ & $0.059 \pm 0.011$ & $0.048\pm 0.012$  & $0.088\pm 0.024$  & $0.1\pm 0.5$  \\
 $d_z$          & $0.002\pm 0.019$ & $-0.02\pm 0.03$   & $0.02\pm 0.04$    & $-0.15\pm 0.07$   & $-0.23\pm 0.13$ \\
\hline
$Q_{zz}$        & $0.03\pm 0.03$   & $0.04\pm 0.05$    & $0.10\pm 0.06$    & $-0.13\pm 0.13$  & $-0.16\pm 0.25$ \\
$Q_{xx}-Q_{yy}$ & $0.018\pm 0.025$ & $0.07\pm 0.04$    & $0.03\pm 0.04$    & $0.18\pm 0.08$   & $0.30\pm 0.17$ \\
$Q_{xy}$        & $-0.016\pm 0.012$ & $0.026\pm 0.019$ & $0.041\pm 0.022$  & $-0.05\pm 0.04$  & $0.11\pm 0.08$ \\
$Q_{xz}$        & $-0.010\pm 0.016$ & $0.017\pm 0.025$ & $0.003\pm 0.029$  & $0.10\pm 0.06$   & $-0.10\pm 0.10$ \\
$Q_{yz}$        & $-0.019\pm 0.016$ & $0.005\pm 0.025$ & $-0.029\pm 0.029$ & $0.09\pm 0.06$   & $0.13\pm 0.10$ \\
\hline
$Q$             & $0.018\pm 0.010$ &  $0.028\pm 0.015$ & $0.05\pm 0.02$    & $0.10\pm 0.03$  & $0.13\pm 0.06$ \\
$Q^{\rm UL}$        & $0.04$           &  $ 0.05$           & $0.08$            & $0.15$          & $0.26$ \\
\hline
\enddata
\end{deluxetable*}

\subsection{Angular power spectrum above 4 EeV}

\begin{figure*}[!ht]
\gridline{\fig{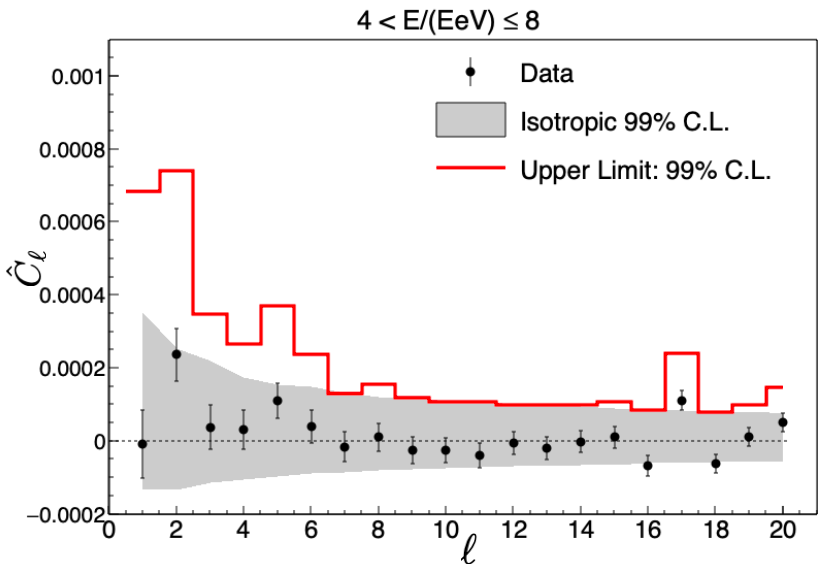}{0.441\textwidth}{(a)}
          \fig{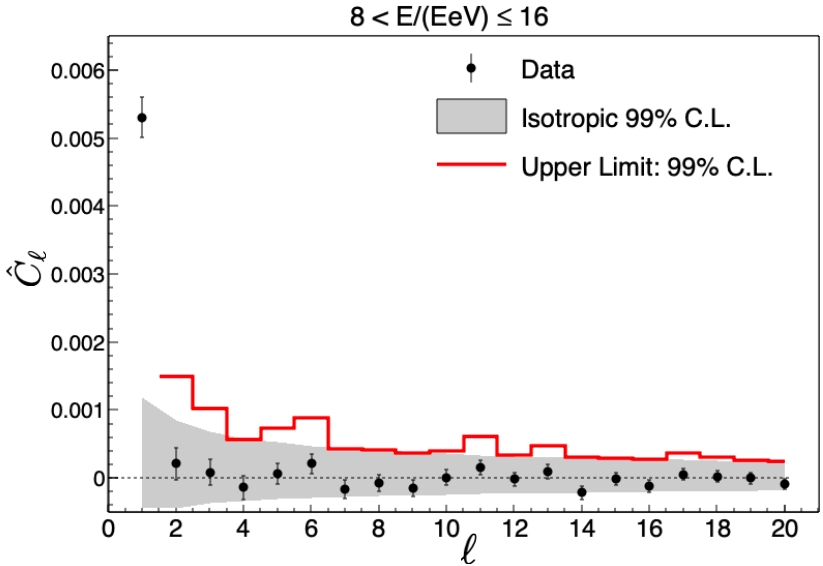}{0.44\textwidth}{(b)}
}
\vspace{-0.4cm}
\gridline{\fig{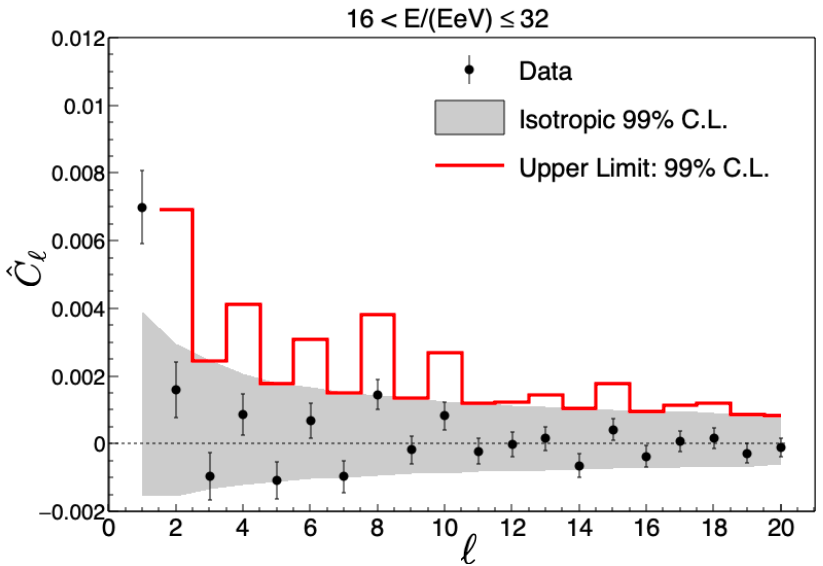}{0.44\textwidth}{(c)}
          \fig{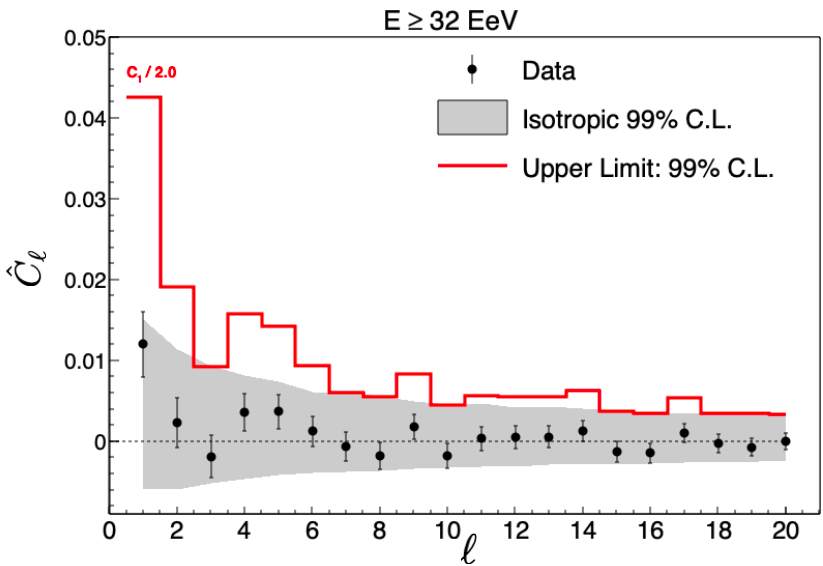}{0.44\textwidth}{(d)}
          }
\vspace{-0.8cm}
\gridline{\fig{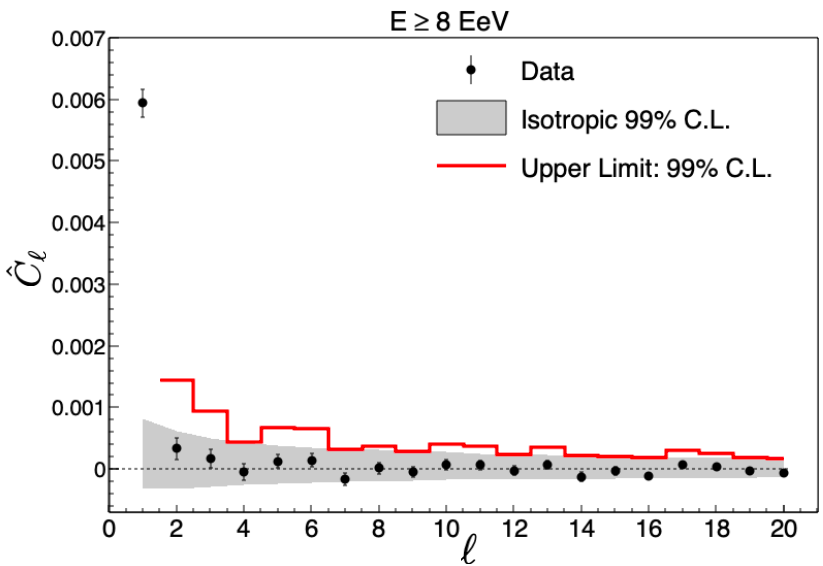}{0.44\textwidth}{(e)}}    
\caption{Angular power spectrum measurements for the (4-8)\,EeV (a), (8-16)\,EeV (b), (16-32)\,EeV (c), $\ge32$\,EeV (d), and $\ge8$\,EeV (e) energy bins. The gray bands correspond to the 99\% CL fluctuations that would result from an isotropic distribution. The red lines correspond to the 99\% CL upper limits. In panel (d), the upper limit $C_{1}^{\rm{UL}}$ has been divided by 2 to maintain the visibility of other features in the plot.\label{Fig:power_spectrum}}

\end{figure*}

The measured power spectra for different energy bins  are presented in Fig.\!\ \ref{Fig:power_spectrum}. The gray band, obtained from Monte Carlo simulations, corresponds to the 99\% CL interval that would result from fluctuations of isotropic distributions. This band was constructed by performing 50,000 isotropic simulations, each containing the same number of events as the data. The $\hat{C}_1$ and $\hat{C}_2$ obtained for all energy bins are consistent with the results from the previous Sections 4.1 and 4.2 using a Fourier analysis. Besides the significant dipolar pattern corresponding to $\hat{C}_1$, the only $\hat{C}_{\ell}$ values that stand above the $99\%$ CL of isotropic fluctuations are  $\hat{C}_{17}$, corresponding to an angular scale of ${\sim}180^{\circ}/\ell \approx 11^{\circ}$, and $\hat{C}_{8}$, corresponding to an angular scale of ${\sim}23^{\circ}$, for the energy bins of $(4,8)$\,EeV and $(16,32)$\,EeV, respectively. After statistical penalization for searches over different multipoles and energy bins (four independent energy bins $\times$ 20 multipole measurements = 80), the probability of these results arising from fluctuations of isotropy are $3.3\%$ and $26.5\%$, respectively. All other $\hat{C}_{\ell}$ values in different energy bins are not significant. 

The red lines indicate the upper limits on multipole amplitudes with $99\%$ CL. They were obtained by using the following approach. For each possible amplitude $C_{\ell}$, the probability of reconstructing $\hat{C}_{\ell}^{\rm{rec}}$ given a true value of $C_{\ell}$,  $\rm{p}$$(\hat{C}_{\ell}^{\rm{rec}},C_{\ell})$, is estimated from simulations.  For this, cosmic-ray skies were simulated by drawing events according to an underlying anisotropic distribution corresponding to $C_{\ell}$, considering an observatory with relative exposure $\rm{W}$. The underlying anisotropic distribution is achieved by generating $a_{\ell m}$ coefficients uniformly distributed on the surface of a $(2\ell+1)$-dimensional hypersphere of radius $\sqrt{(2\ell + 1)C_{\ell}}$. The reconstructed $\hat{C}_{\ell}^{\rm{rec}}$  for each simulated cosmic-ray sky are obtained using Eq.~\ref{eqn:Cl}. The amplitude $C_{\ell}^{\rm{UL}}$ such that $\int_{\hat{C}_{\ell, \rm{data}}}^{\infty} d\hat{C}^{\rm{rec}}_{\ell}\rm{p}$$(\hat{C}_{\ell}^{\rm{rec}},C_{\ell}^{\rm{UL}} ) = \rm{CL}$ is the relevant upper limit\footnote{Since this procedure can lead to upper limits tighter than the upper bounds obtained from isotropy with $99\%$ CL, $\hat{C}_{\ell,\rm{iso}}^{99}$, when the measured values of $\hat{C}_{\ell,\rm{data}}$ are smaller than the expected average for isotropy, the upper limits presented are defined as $\rm max$$(C_{\ell}^{\rm{UL}}, \hat{C}_{\ell,\rm{iso}}^{99})$ in order to cope with this undesired behavior.}.  
\subsection{Modulation in R.A. down to 0.03 EeV}

\begin{figure*}[!ht]
\gridline{\fig{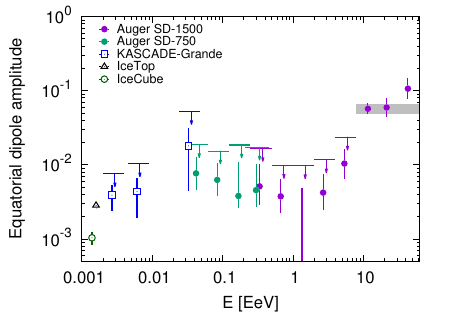}{0.45\textwidth}{(a)}
          \fig{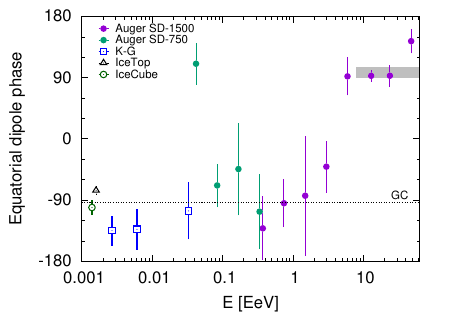}{0.45\textwidth}{(b)}
}
\caption{Equatorial dipole amplitude (a) and phase (b) for the energy bins where the dataset from the SD-1500 array (purple circles) or the SD-750 array (green circles) are used. The $99\%$ CL upper limits for the energy bins in which the obtained amplitude has a P($\ge r_1^\alpha$)$>1\%$ are shown. Results from the IceCube and KASCADE-Grande Collaborations are also included for comparison \citep{IceCube:2011uwn, IceCube:2016biq, Apel:2019afz}. 
    \label{fig:LSA_dipole_lowE}}
\end{figure*}

The modulation in R.A. is studied for energies below 4\,EeV using the data of the SD-1500 and SD-750 arrays, with the East-West method or a Fourier analysis, as described in Section~3.4. The results of these analyses, the equatorial amplitude ($d_\perp$), phase ($\alpha_d$) and isotropic probability ($P(\ge r_1^\alpha)$), are presented in Table~\ref{tab:LSA_lowE}, where it is indicated which method was used in each energy bin. The $99\%$ CL upper limit, $d_\perp^{\rm UL}$ is reported. These upper limits are derived from the distribution for a dipolar anisotropy of unknown amplitude, marginalized over the dipole phase, as described in \citep{2020ApJ...891..142A}. In Fig.~\ref{fig:LSA_dipole_lowE}, the results are shown, also including those obtained at lower energies by the IceCube and KASCADE-Grande Collaborations \citep{IceCube:2011uwn, IceCube:2016biq, Apel:2019afz}.

Due to the small amplitudes ($<1\%$) and the current amount of statistics, the results listed in Table~\ref{tab:LSA_lowE} are not statistically significant ($P(\ge r_1^\alpha)>1\%$) and it is not yet possible to draw firm conclusions. Nonetheless, it is suggestive that the amplitudes of the equatorial dipole increase as a function of energy, from below $1\%$ to above $10\%$, and that the phases shift from close to the Galactic Center to the opposite direction. These changes in amplitude and phase could suggest a transition of the origin of the anisotropies from Galactic cosmic rays to extragalactic ones. Another explanation for these changes could be the impact of the Galactic magnetic field on an extragalactic flux, as discussed in Section 5.

\begin{deluxetable*}{cccccccc}[!ht]
\tablecaption{{Results for the large scale analysis in R.A..  For each energy bin, the number of events, $N$, the equatorial component of the amplitude, $d_\perp$, the R.A. of the dipole direction, $\alpha_d$, the probability of getting a larger amplitude from fluctuations of an isotropic distribution, $P(\ge r_1^\alpha)$, and the $99\%$ CL upper limit, $d_\perp^{\rm UL}$, are presented.}}
 \label{tab:LSA_lowE}
\tablehead{\colhead{ }
	 & \colhead{ } & \colhead{$E$ [${\rm EeV}$]} & \colhead{$N$} & \colhead{$d_\perp (\%)$} &	\colhead{$\alpha_d$ [$^\circ$]} & \colhead{$P(\ge r_1^\alpha)$} & \colhead{$d_\perp^{\rm UL} (\%)$}}
\startdata
$\boldsymbol{{\rm SD}750}$ & ${\rm East-West}$ & 1/32-1/16 & 1,811,897 & $0.8^{+0.5}_{-0.3}$ & $110 \pm 31$ & 0.22 & 1.9 \\      
    & & 1/16-1/8 & 1,843,507 & $0.6^{+0.4}_{-0.2}$ & $-69 \pm 32$ &  0.23 & 1.5 \\       
    & & 1/8-1/4 & 607,690 & $0.4^{+0.7}_{-0.1}$ & $-44 \pm 68$ &  0.79 & 1.8 \\
& ${\rm Fourier}$ & 0.25-0.5 & 135,182  & $0.5^{+0.6}_{-0.2}$ & $-107 \pm 55$ & 0.65 & 1.7\\ 
    \hline 
$\boldsymbol{{\rm SD}1500}$ & ${\rm East-West}$ & 0.25-0.5 &  930,942 & $0.5^{+0.5}_{-0.2}$ & $-132 \pm 47$ & 0.51 & 1.7\\     
     &  & 0.5 - 1 & 3,049,342 & $0.4^{+0.3}_{-0.2}$  & $-95 \pm 35$ & 0.28 & 1.0\\ 
         &  & 1-2 & 1,639,139 & $0.1^{+0.4}_{-0.1}$     & $-84 \pm 88$ & 0.93 & 1.0 \\
& ${\rm Fourier}$ & 2-4 & 380,491 & $0.4^{+0.3}_{-0.2}$ & $-41 \pm 38$ & 0.36 & 1.2 \\ 
\enddata
\end{deluxetable*}

\section{Model predictions and discussion}

Possible explanations of the origin of the dipole measured above $8\,$EeV assume an inhomogeneous distribution of sources, probably following the large-scale structure distribution in our neighborhood \citep{Harari:2013pea,Harari:2015hba,Tinyakov:2014fwa, diMatteo:2017dtg, Globus:2017fym, Ding:2021emg, Allard:2021ioh, Bister:2023icg}, or that ultra-high-energy cosmic rays originate from a dominant local source and propagate diffusively through intergalactic magnetic fields \citep{1980JPhG....6.1561G,Berezinskii:1990ap,Mollerach:2019wne}. The resulting dipole amplitude and direction as a function of energy depend on the source distribution, intervening magnetic fields, as well as on the spectrum and mass composition of the UHECRs. If the sources were distributed like galaxies and if deflections in the Galactic magnetic field could be neglected, a dipolar cosmic-ray anisotropy would be expected in a direction close to that of the dipole associated with the galaxies themselves  \citep{Erdogdu:2005wi}. On average, a smaller dipolar anisotropy is expected for larger source densities or for more isotropically distributed sources. The amplitude is predicted to grow for increasing energies as a consequence of the shrinking  of the horizon due to interactions with the radiation backgrounds (resulting in a larger contribution of the nearby non-homogeneously distributed sources). Moreover, the Galactic magnetic field modifies the direction and amplitude of the dipolar anisotropy observed at Earth with respect to that of the particles reaching the halo \citep{Harari:2010wq,2018ApJ...868....4A}. These effects depend on the rigidity of the incoming particles and usually lead to a decrease in the dipolar amplitude and also affects the higher multipoles. In particular, for sources following the distribution of galaxies and considering a mean rigidity of the cosmic rays above $8\,$EeV compatible with the measured mass composition, the direction of the expected dipole would be transformed at Earth after the deflections in the Galactic magnetic field to a direction closer to the observed dipole direction \citep{Science2017}.

The energy spectrum and mass composition data measured with the Pierre Auger Observatory above $6\times 10^{17}\,$eV can be described by a model with two populations of sources, dominating the flux above and below few EeV, respectively \citep{PierreAuger:2022atd}. It has been shown that the data can be well reproduced if the high-energy extragalactic population emits a mixture of heavy and intermediate mass nuclei with a very hard spectrum and a small rigidity cutoff, while the low-energy component is a mixture of light and intermediate mass nuclei with a soft spectrum. 

In this work, the expected dipole amplitude as a function of energy and the expected directions at Earth are computed for sources emitting cosmic rays according to the best fitting model for the high-energy population in \citep{PierreAuger:2022atd} and distributed following the infrared-detected galaxies in the Two Micron All-Sky Survey (2MASS) catalog \citep{2MASS:2006qir} described in \citep{PierreAuger:2022axr} (with distances from the HyperLEDA database \citep{Makarov:2014txa}). The expectation for the dipole amplitude is calculated in the four energy bins above $4\,$EeV for a population of equal-luminosity sources with number density $10^{-5}\,{\rm Mpc}^{-3}$ and $10^{-4}\,{\rm Mpc}^{-3}$ selected randomly from a volume limited sample of the infrared (IR) catalog up to $120\,$Mpc (and from a uniform distribution for larger distances and in the region of the Galactic plane mask). The sources are considered to emit a mixed composition of particles with spectrum $E^{2}$ with a rigidity cutoff $R_{\rm cut}= 10^{18.15}\,$eV \citep{PierreAuger:2022atd}. The fraction of the total source emissivity above $10^{17.8}\,$eV in each mass group is $(I_{\rm H}, I_{\rm He}, I_{\rm N}, I_{\rm Si}, I_{\rm Fe}) = (0., 0.236, 0.721, 0.013, 0.031)$ \citep{PierreAuger:2022atd}. The particles were propagated taking into account the interactions with the extragalactic-background radiation as described by the Gilmore et al. model \citep{Gilmore:2011ks} using the SimProp propagation code \citep{Aloisio:2017iyh}. The arrival direction of the particles at the halo was mapped to the arrival direction at Earth taking into account the deflection in the regular Galactic magnetic field according to the rigidity of the particles and considering for reference the field model from \citep{Jansson:2012pc}. The dipole direction before magnetic deflections for the whole IR catalog up to $120\,$Mpc is ($l, b$) = ($255^\circ,50^\circ$).

\begin{figure}[!ht]
    \centering \includegraphics[width=.49\textwidth]{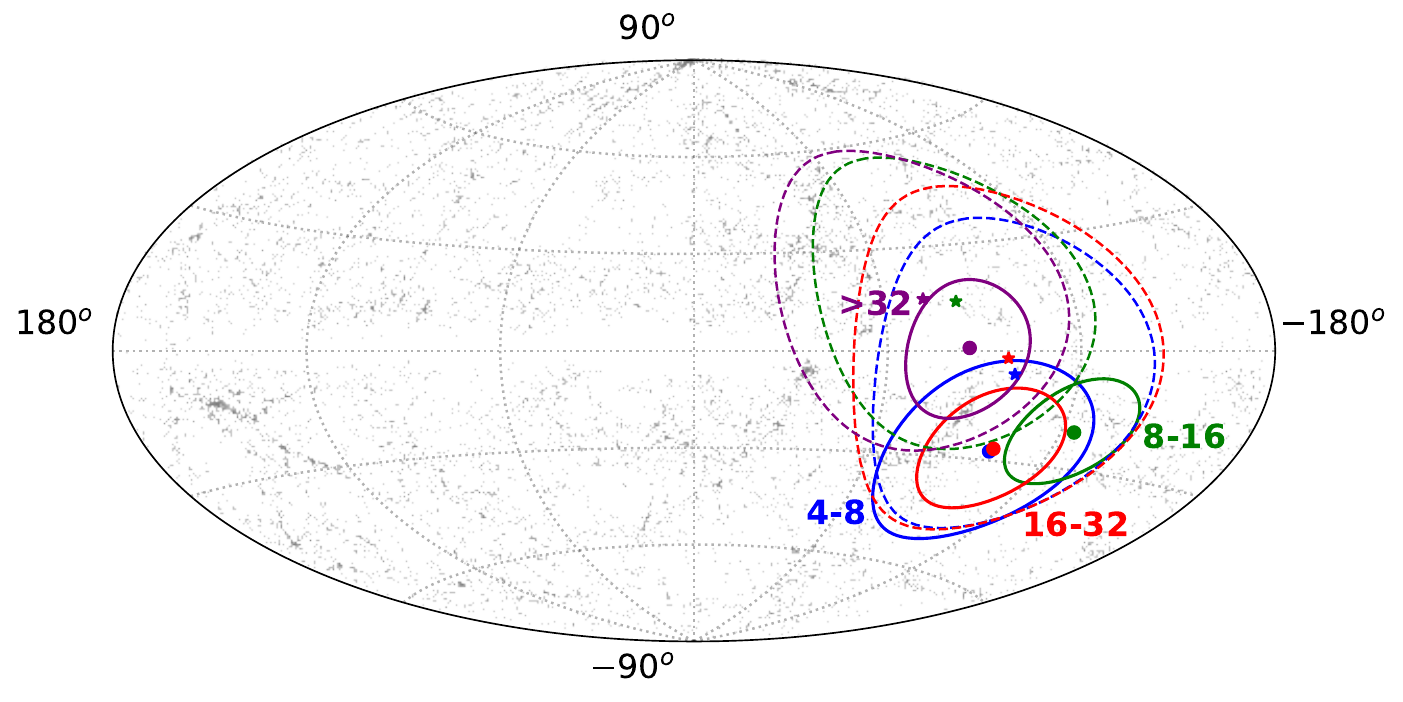}
    \caption{Map in Galactic coordinates showing the predictions for the direction of the mean dipole (star symbols) and the $68\%$ CL contour regions (dashed lines) obtained for $10^3$ realizations of the source distribution for a density of $10^{-4}\,{\rm Mpc}^{-3}$ and for each energy bin above $4\,$EeV. This is compared to what obtained in data (continuous lines). The gray dots represent the location of the galaxies in the IR catalog within $120\,$Mpc.}.
    \label{fig:pred_dir}
\end{figure}

\begin{figure*}[!ht]
\gridline{\fig{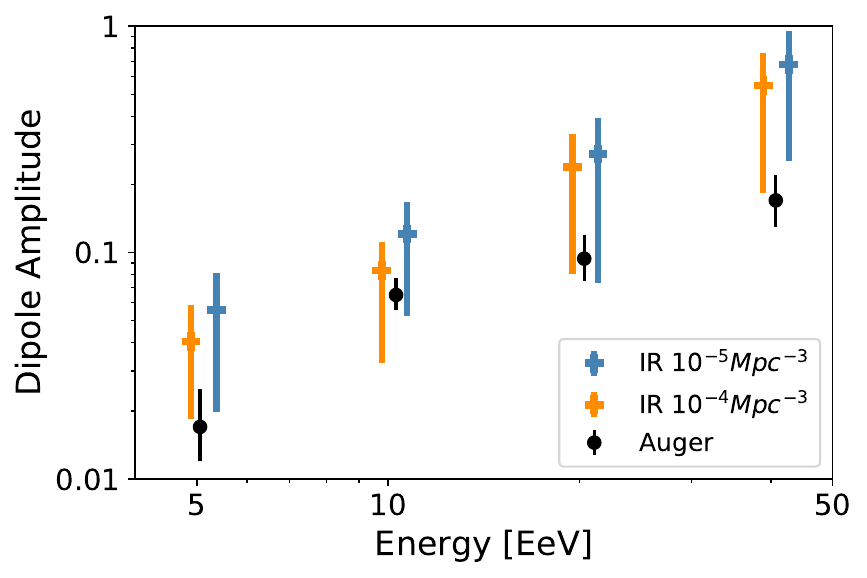}{0.45\textwidth}{(a)}
          \fig{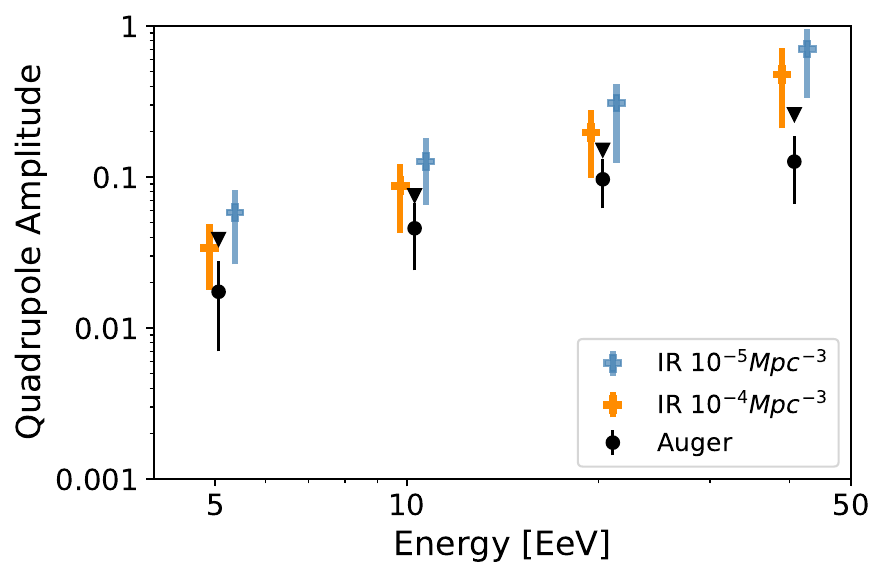}{0.45\textwidth}{(b)}}
\caption{(a) Median and $68\%$ CL range of the dipole amplitudes for the two source densities considered, $10^{-4}\,{\rm Mpc}^{-3}$ (orange) and $10^{-5}\,{\rm Mpc}^{-3}$ (blue). (b) Expected values of the average quadrupole amplitude, $Q$, for the same model of the high-energy population of sources. In both plots the results from data are shown (black dots) and for the average quadrupole amplitude the $99\%$ CL upper limits are included (black triangles). The four energy ranges are (4-8, 8-16, 16-32, $\ge 32$)\,EeV.\label{fig:dipquadampHE}}
\end{figure*}

In Fig.~\ref{fig:pred_dir} the direction of the mean dipole of the simulations and the $68\%$ CL sky region obtained for $10^3$ realizations of the source distribution for a density of $10^{-4}\,{\rm Mpc}^{-3}$ are presented. The results for a source density of $10^{-5}\,{\rm Mpc}^{-3}$ are similar, but the contour regions are larger due to the larger fluctuations in the source distribution.\footnote{Note that back-tracking studies in the UF23 GMF models \citep{Unger:2023lob}, including also a turbulent component, show that the direction of the dipole of CRs with $E/Z = 1\,$EeV having a dipolar distribution with maximum in the direction of the 2MRS dipole outside the Galaxy, have after traversing the Galactic magnetic field a median angular separation of $17.5^\circ$ (or $15.1^\circ$ for $E/Z = 4\,$EeV) with respect to the location obtained with the JF12 model. This uncertainty coming from the GMF models is smaller than the uncertainty arising from the source distribution.} In the left panel of Fig.~\ref{fig:dipquadampHE} the median and $68\%$ CL range of the dipole amplitudes are shown for the two source densities considered, also including for comparison the results from data. The expected values of the average quadrupole amplitude, $Q$, for the same models of the high-energy source population, are displayed in the right panel of Fig.~\ref{fig:dipquadampHE}, together with the results from data and the $99\%$ CL upper limits obtained for $Q$. The dipolar and quadrupolar anisotropies for both source densities are compatible with the experimental results within the uncertainties, although for the smallest source density the quadrupole prediction is in slight tension, in particular for the highest energy bin. Possible ways to reduce the quadrupole prediction, besides increasing the source density considered, would be to invoke strong turbulent Galactic and/or extragalactic magnetic field deflections to smooth out the arrival direction maps \citep{Allard:2021ioh,Bister:2023icg}.

For energies below the ankle, the results for the low-energy component of the combined fit of the energy spectrum and mass composition measured with the Pierre Auger Observatory can be described by two different scenarios \citep{PierreAuger:2022atd}. The first one consists of a Galactic contribution of nitrogen and an extragalactic contribution of protons (that could be produced by interactions of nuclei from the high-energy population in the environment of the sources). The second one consists of an extragalactic contribution of mixed composition (proton, helium and nitrogen). In both scenarios, there is also a high-energy extragalactic component with mixed composition, as considered before. 

The dipolar anisotropy of the nitrogen Galactic component of the first scenario is expected to point close to the Galactic center, in agreement with the measured right ascension phase at low energies, although its amplitude may be close to the present upper bounds at EeV energies \citep{PierreAuger:2012gro}.

Regarding the second scenario, the arrival direction distribution of CRs from the lower-energy extragalactic component is expected to be more isotropic, since at those energies cosmic rays can reach the Earth from distances larger than $1\,$Gpc and the distribution of matter in the Universe at very large scales looks quite uniform. However, even in the case of a perfectly uniform distribution of cosmic rays in some reference frame outside the Galaxy, the distribution of arrival directions at Earth is not expected to be isotropic. In particular, the relative motion of the observer with respect to the isotropic reference frame gives rise to an anisotropy that is expected to be dipolar before traversing the Galactic magnetic field, the so-called Compton-Getting effect. If the isotropic reference frame is taken as that of the Cosmic Microwave Background (CMB), this dipolar anisotropy amplitude has been estimated to be of order $6\times 10^{-3}$ \citep{Kachelriess:2006aq}, pointing in the direction of the CMB dipole, $(l, b) = (264^\circ, 48^\circ)$ \citep{2020A&A...641A...1P}. This amplitude is an order of magnitude smaller than the measured anisotropies at the highest energies, and thus represents a subdominant effect above the ankle, but it is of the order of the anisotropies measured below the ankle. The propagation in the Galactic magnetic field distorts the dipolar pattern, changing the amplitude and direction of the dipole observed at Earth, which can have a notable effect on the predictions for the right ascension phase  \citep{Mollerach:2022aji}. In particular, the dipole direction may shift below few EeV towards the inner spiral arm direction, at $(l, b) \simeq (-100^\circ , 0^\circ )$, which can be relevant to explain the observed change in phase.  Also, as a consequence of the Galaxy rotation the Galactic magnetic field has a small electric component which leads to a direction dependent CR acceleration, possibly affecting the anisotropies at a level comparable to the Compton-Getting effect \citep{Mollerach:2022aji}.

Clearly more precise measurements of the anisotropies in the 0.1\,EeV to few EeV energy range, possibly discriminated by mass composition, as well as a better knowledge of the Galactic magnetic field, will be very useful to shed light on the origin of the CRs below the ankle and on the details of the Galactic to extragalactic transition. 

\section{Conclusions}

The results of the anisotropy searches using the arrival directions of UHECRs detected in the Phase 1 of operation of the Pierre Auger Observatory, corresponding to the 19 years of data gathered before the implementation of the AugerPrime upgrade, were presented. The significance of the established equatorial dipole for the cumulative energy bin above $8\,$EeV is now $6.8\sigma$, and that for the bin 8-$16\,$EeV is $5.7\sigma$, surpassing the discovery level. For the $\ge8\,$EeV cumulative energy bin, the statistics are such that we can divide the dataset in time-ordered bins and we find no time variation of the equatorial dipole modulation with an upper limit on the long-term rate of change of 0.003 per year at the 95\% CL. If a quadrupolar distribution is allowed in the flux parametrization, the obtained quadrupolar moments are not statistically significant. 

The dipole amplitudes increase with energy above 4\,EeV. Model predictions were presented for sources following the distribution of galaxies, assuming a source density of $10^{-5}\,{\rm Mpc}^{-3}$ or $10^{-4}\,{\rm Mpc}^{-3}$, and emitting  according to the model for the high-energy population from our combined fit of spectrum and composition \citep{PierreAuger:2022atd}. The predictions for the dipole amplitude and direction were shown, as well as for the quadrupole amplitude, which are consistent with data within their uncertainties, although some tension with the small observed quadrupole amplitudes seems to be present, in particular for the lowest source density considered and the highest energy bin.

By studying the R.A. distribution of the events, the equatorial component of the dipole down to $0.03\,$EeV was computed. For energies below 4\,EeV the amplitudes are below $1\%$, compatible with isotropic expectations. However, in most of the energy bins the phase points are consistently close to the Galactic Center phase, and nearly opposite to the Galactic Center phase at energies above 4\,EeV. This could be due to the observed anisotropy having a predominant Galactic origin below $1\,$EeV and a predominant extragalactic origin above few EeV. Alternatively, this could  be caused by the effects of the Galactic magnetic field on an extragalactic flux.

The encouraging prospects of obtaining mass-composition estimators on an event-by-event basis with AugerPrime, and improved mass estimators with Phase 1 data, will give more information about these anisotropies. An analysis including these estimators to study the different dipole amplitudes obtained  separating the ``lighter'' and ``heavier'' events with the present dataset is forthcoming.\\

\vspace{-1ex}
\footnotesize
\section*{Acknowledgments}

\begin{sloppypar}
The successful installation, commissioning, and operation of the Pierre
Auger Observatory would not have been possible without the strong
commitment and effort from the technical and administrative staff in
Malarg\"ue. We are very grateful to the following agencies and
organizations for financial support:
\end{sloppypar}

\begin{sloppypar}
Argentina -- Comisi\'on Nacional de Energ\'\i{}a At\'omica; Agencia Nacional de
Promoci\'on Cient\'\i{}fica y Tecnol\'ogica (ANPCyT); Consejo Nacional de
Investigaciones Cient\'\i{}ficas y T\'ecnicas (CONICET); Gobierno de la
Provincia de Mendoza; Municipalidad de Malarg\"ue; NDM Holdings and Valle
Las Le\~nas; in gratitude for their continuing cooperation over land
access; Australia -- the Australian Research Council; Belgium -- Fonds
de la Recherche Scientifique (FNRS); Research Foundation Flanders (FWO),
Marie Curie Action of the European Union Grant No.~101107047; Brazil --
Conselho Nacional de Desenvolvimento Cient\'\i{}fico e Tecnol\'ogico (CNPq);
Financiadora de Estudos e Projetos (FINEP); Funda\c{c}\~ao de Amparo \`a
Pesquisa do Estado de Rio de Janeiro (FAPERJ); S\~ao Paulo Research
Foundation (FAPESP) Grants No.~2019/10151-2, No.~2010/07359-6 and
No.~1999/05404-3; Minist\'erio da Ci\^encia, Tecnologia, Inova\c{c}\~oes e
Comunica\c{c}\~oes (MCTIC); Czech Republic -- GACR 24-13049S, CAS LQ100102401,
MEYS LM2023032, CZ.02.1.01/0.0/0.0/16{\textunderscore}013/0001402,
CZ.02.1.01/0.0/0.0/18{\textunderscore}046/0016010 and
CZ.02.1.01/0.0/0.0/17{\textunderscore}049/0008422 and CZ.02.01.01/00/22{\textunderscore}008/0004632;
France -- Centre de Calcul IN2P3/CNRS; Centre National de la Recherche
Scientifique (CNRS); Conseil R\'egional Ile-de-France; D\'epartement
Physique Nucl\'eaire et Corpusculaire (PNC-IN2P3/CNRS); D\'epartement
Sciences de l'Univers (SDU-INSU/CNRS); Institut Lagrange de Paris (ILP)
Grant No.~LABEX ANR-10-LABX-63 within the Investissements d'Avenir
Programme Grant No.~ANR-11-IDEX-0004-02; Germany -- Bundesministerium
f\"ur Bildung und Forschung (BMBF); Deutsche Forschungsgemeinschaft (DFG);
Finanzministerium Baden-W\"urttemberg; Helmholtz Alliance for
Astroparticle Physics (HAP); Helmholtz-Gemeinschaft Deutscher
Forschungszentren (HGF); Ministerium f\"ur Kultur und Wissenschaft des
Landes Nordrhein-Westfalen; Ministerium f\"ur Wissenschaft, Forschung und
Kunst des Landes Baden-W\"urttemberg; Italy -- Istituto Nazionale di
Fisica Nucleare (INFN); Istituto Nazionale di Astrofisica (INAF);
Ministero dell'Universit\`a e della Ricerca (MUR); CETEMPS Center of
Excellence; Ministero degli Affari Esteri (MAE), ICSC Centro Nazionale
di Ricerca in High Performance Computing, Big Data and Quantum
Computing, funded by European Union NextGenerationEU, reference code
CN{\textunderscore}00000013; M\'exico -- Consejo Nacional de Ciencia y Tecnolog\'\i{}a
(CONACYT) No.~167733; Universidad Nacional Aut\'onoma de M\'exico (UNAM);
PAPIIT DGAPA-UNAM; The Netherlands -- Ministry of Education, Culture and
Science; Netherlands Organisation for Scientific Research (NWO); Dutch
national e-infrastructure with the support of SURF Cooperative; Poland
-- Ministry of Education and Science, grants No.~DIR/WK/2018/11 and
2022/WK/12; National Science Centre, grants No.~2016/22/M/ST9/00198,
2016/23/B/ST9/01635, 2020/39/B/ST9/01398, and 2022/45/B/ST9/02163;
Portugal -- Portuguese national funds and FEDER funds within Programa
Operacional Factores de Competitividade through Funda\c{c}\~ao para a Ci\^encia
e a Tecnologia (COMPETE); Romania -- Ministry of Research, Innovation
and Digitization, CNCS-UEFISCDI, contract no.~30N/2023 under Romanian
National Core Program LAPLAS VII, grant no.~PN 23 21 01 02 and project
number PN-III-P1-1.1-TE-2021-0924/TE57/2022, within PNCDI III; Slovenia
-- Slovenian Research Agency, grants P1-0031, P1-0385, I0-0033, N1-0111;
Spain -- Ministerio de Ciencia e Innovaci\'on/Agencia Estatal de
Investigaci\'on (PID2019-105544GB-I00, PID2022-140510NB-I00 and
RYC2019-027017-I), Xunta de Galicia (CIGUS Network of Research Centers,
Consolidaci\'on 2021 GRC GI-2033, ED431C-2021/22 and ED431F-2022/15),
Junta de Andaluc\'\i{}a (SOMM17/6104/UGR and P18-FR-4314), and the European
Union (Marie Sklodowska-Curie 101065027 and ERDF); USA -- Department of
Energy, Contracts No.~DE-AC02-07CH11359, No.~DE-FR02-04ER41300,
No.~DE-FG02-99ER41107 and No.~DE-SC0011689; National Science Foundation,
Grant No.~0450696, and NSF-2013199; The Grainger Foundation; Marie
Curie-IRSES/EPLANET; European Particle Physics Latin American Network;
and UNESCO.
\end{sloppypar}

\bibliography{biblio}{}

\begin{thebibliography}{}
\expandafter\ifx\csname natexlab\endcsname\relax\def\natexlab#1{#1}\fi
\providecommand{\url}[1]{\href{#1}{#1}}
\providecommand{\dodoi}[1]{doi:~\href{http://doi.org/#1}{\nolinkurl{#1}}}
\providecommand{\doeprint}[1]{\href{http://ascl.net/#1}{\nolinkurl{http://ascl.net/#1}}}
\providecommand{\doarXiv}[1]{\href{https://arxiv.org/abs/#1}{\nolinkurl{https://arxiv.org/abs/#1}}}

\bibitem[{Allard {et~al.}(2022)Allard, Aublin, Baret, \&
  Parizot}]{Allard:2021ioh}
Allard, D., Aublin, J., Baret, B., \& Parizot, E. 2022, Astron. Astrophys.,
  664, A120, \dodoi{10.1051/0004-6361/202142491}

\bibitem[{Aloisio {et~al.}(2017)Aloisio, Boncioli, Di~Matteo, Grillo, Petrera,
  \& Salamida}]{Aloisio:2017iyh}
Aloisio, R., Boncioli, D., Di~Matteo, A., {et~al.} 2017, JCAP, 11, 009,
  \dodoi{10.1088/1475-7516/2017/11/009}

\bibitem[{Berezinskii {et~al.}(1990)Berezinskii, Grigor'eva, \&
  Dogiel}]{Berezinskii:1990ap}
Berezinskii, V.~S., Grigor'eva, S.~I., \& Dogiel, V.~A. 1990, Astron.
  Astrophys., 232, 582

\bibitem[{Bister \& Farrar(2024)}]{Bister:2023icg}
Bister, T., \& Farrar, G.~R. 2024, Astrophys. J., 966, 71,
  \dodoi{10.3847/1538-4357/ad2f3f}

\bibitem[{Bonino {et~al.}(2011)Bonino, Alekseenko, Deligny, Ghia, Grigat,
  Letessier-Selvon, Lyberis, Mollerach, Over, \& Roulet}]{Bonino:2011nx}
Bonino, R., Alekseenko, V.~V., Deligny, O., {et~al.} 2011, Astrophys. J., 738,
  67, \dodoi{10.1088/0004-637X/738/1/67}

\bibitem[{{Deligny} {et~al.}(2004){Deligny}, {Armengaud}, {Beau}, {Da Silva},
  {Hamilton}, {Lachaud}, {Letessier-Selvon}, \& {Revenu}}]{Deligny2004}
{Deligny}, O., {Armengaud}, E., {Beau}, T., {et~al.} 2004, \jcap, 008.
\newblock \doarXiv{astro-ph/0404253}

\bibitem[{di~Matteo \& Tinyakov(2018)}]{diMatteo:2017dtg}
di~Matteo, A., \& Tinyakov, P. 2018, Mon. Not. Roy. Astron. Soc., 476, 715,
  \dodoi{10.1093/mnras/sty277/4835522}

\bibitem[{Ding {et~al.}(2021)Ding, Globus, \& Farrar}]{Ding:2021emg}
Ding, C., Globus, N., \& Farrar, G.~R. 2021, Astrophys. J. Lett., 913, L13,
  \dodoi{10.3847/2041-8213/abf11e}

\bibitem[{Erdogdu {et~al.}(2006)}]{Erdogdu:2005wi}
Erdogdu, P., {et~al.} 2006, Mon. Not. Roy. Astron. Soc., 368, 1515,
  \dodoi{10.1111/j.1365-2966.2006.10243.x}

\bibitem[{{Giler} {et~al.}(1980){Giler}, {Wdowczyk}, \&
  {Wolfendale}}]{1980JPhG....6.1561G}
{Giler}, M., {Wdowczyk}, J., \& {Wolfendale}, A.~W. 1980, Journal of Physics G
  Nuclear Physics, 6, 1561, \dodoi{10.1088/0305-4616/6/12/011}

\bibitem[{Gilmore {et~al.}(2012)Gilmore, Somerville, Primack, \&
  Dominguez}]{Gilmore:2011ks}
Gilmore, R.~C., Somerville, R.~S., Primack, J.~R., \& Dominguez, A. 2012, Mon.
  Not. Roy. Astron. Soc., 422, 3189, \dodoi{10.1111/j.1365-2966.2012.20841.x}

\bibitem[{Globus \& Piran(2017)}]{Globus:2017fym}
Globus, N., \& Piran, T. 2017, Astrophys. J. Lett., 850, L25,
  \dodoi{10.3847/2041-8213/aa991b}

\bibitem[{{Greisen}(1966)}]{1966PhRvL..16..748G}
{Greisen}, K. 1966, \prl, 16, 748, \dodoi{10.1103/PhysRevLett.16.748}

\bibitem[{Harari {et~al.}(2010)Harari, Mollerach, \& Roulet}]{Harari:2010wq}
Harari, D., Mollerach, S., \& Roulet, E. 2010, JCAP, 11, 033,
  \dodoi{10.1088/1475-7516/2010/11/033}

\bibitem[{Harari {et~al.}(2014)Harari, Mollerach, \& Roulet}]{Harari:2013pea}
---. 2014, Phys. Rev. D, 89, 123001, \dodoi{10.1103/PhysRevD.89.123001}

\bibitem[{Harari {et~al.}(2015)Harari, Mollerach, \& Roulet}]{Harari:2015hba}
---. 2015, Phys. Rev. D, 92, 063014, \dodoi{10.1103/PhysRevD.92.063014}

\bibitem[{{IceCube Collaboration}(2012)}]{IceCube:2011uwn}
{IceCube Collaboration}. 2012, Astrophys. J., 746, 33,
  \dodoi{10.1088/0004-637X/746/1/33}

\bibitem[{{IceCube Collaboration}(2016)}]{IceCube:2016biq}
---. 2016, Astrophys. J., 826, 220, \dodoi{10.3847/0004-637X/826/2/220}

\bibitem[{Jansson \& Farrar(2012)}]{Jansson:2012pc}
Jansson, R., \& Farrar, G.~R. 2012, Astrophys. J., 757, 14,
  \dodoi{10.1088/0004-637X/757/1/14}

\bibitem[{Kachelriess \& Serpico(2006)}]{Kachelriess:2006aq}
Kachelriess, M., \& Serpico, P.~D. 2006, Phys. Lett. B, 640, 225,
  \dodoi{10.1016/j.physletb.2006.08.006}

\bibitem[{{KASCADE-Grande Collaboration}(2019)}]{Apel:2019afz}
{KASCADE-Grande Collaboration}. 2019, Astrophys. J., 870, 91,
  \dodoi{10.3847/1538-4357/aaf1ca}

\bibitem[{Linsley(1975)}]{Linsley:1975kp}
Linsley, J. 1975, Phys. Rev. Lett., 34, 1530,
  \dodoi{10.1103/PhysRevLett.34.1530}

\bibitem[{Makarov {et~al.}(2014)Makarov, Prugniel, Terekhova, Courtois, \&
  Vauglin}]{Makarov:2014txa}
Makarov, D., Prugniel, P., Terekhova, N., Courtois, H., \& Vauglin, I. 2014,
  Astron. Astrophys., 570, A13, \dodoi{10.1051/0004-6361/201423496}

\bibitem[{Mollerach \& Roulet(2019)}]{Mollerach:2019wne}
Mollerach, S., \& Roulet, E. 2019, Phys. Rev. D, 99, 103010,
  \dodoi{10.1103/PhysRevD.99.103010}

\bibitem[{Mollerach {et~al.}(2022)Mollerach, Roulet, \&
  Taborda}]{Mollerach:2022aji}
Mollerach, S., Roulet, E., \& Taborda, O. 2022, JCAP, 12, 021,
  \dodoi{10.1088/1475-7516/2022/12/021}

\bibitem[{{{Pierre Auger and Telescope Array
  Collaborations}}(2014)}]{TelescopeArray:2014ahm}
{{Pierre Auger and Telescope Array Collaborations}}. 2014, Astrophys. J., 794,
  172, \dodoi{10.1088/0004-637X/794/2/172}

\bibitem[{{{Pierre Auger and Telescope Array
  Collaborations}}(2023)}]{PierreAuger:2023mvf}
---. 2023, PoS, ICRC2023, 521, \dodoi{10.22323/1.444.0521}

\bibitem[{{Pierre Auger Collaboration}(2011)}]{PierreAuger:2011yxe}
{Pierre Auger Collaboration}. 2011, JCAP, 11, 022,
  \dodoi{10.1088/1475-7516/2011/11/022}

\bibitem[{{Pierre Auger Collaboration}(2012)}]{PierreAuger:2012gro}
---. 2012, Astrophys. J. Lett., 762, L13, \dodoi{10.1088/2041-8205/762/1/L13}

\bibitem[{{Pierre Auger Collaboration}(2015{\natexlab{a}})}]{AugerNIM2015}
---. 2015{\natexlab{a}}, NIM A, 798, 172,
  \dodoi{https://doi.org/10.1016/j.nima.2015.06.058}

\bibitem[{{Pierre Auger
  Collaboration}(2015{\natexlab{b}})}]{PierreAuger:2014ati}
---. 2015{\natexlab{b}}, Astrophys. J., 802, 111,
  \dodoi{10.1088/0004-637X/802/2/111}

\bibitem[{{Pierre Auger Collaboration}(2017{\natexlab{a}})}]{Science2017}
---. 2017{\natexlab{a}}, Science, 357, 1266, \dodoi{10.1126/science.aan4338}

\bibitem[{{Pierre Auger
  Collaboration}(2017{\natexlab{b}})}]{PierreAuger:2016gkp}
---. 2017{\natexlab{b}}, JCAP, 06, 026, \dodoi{10.1088/1475-7516/2017/06/026}

\bibitem[{{Pierre Auger
  Collaboration}(2017{\natexlab{c}})}]{PierreAuger:2017vtr}
---. 2017{\natexlab{c}}, JINST, 12, P02006,
  \dodoi{10.1088/1748-0221/12/02/P02006}

\bibitem[{{Pierre Auger Collaboration}(2018)}]{2018ApJ...868....4A}
---. 2018, \apj, 868, 4, \dodoi{10.3847/1538-4357/aae689}

\bibitem[{{Pierre Auger
  Collaboration}(2020{\natexlab{a}})}]{2020ApJ...891..142A}
---. 2020{\natexlab{a}}, \apj, 891, 142, \dodoi{10.3847/1538-4357/ab7236}

\bibitem[{{Pierre Auger
  Collaboration}(2020{\natexlab{b}})}]{2020PhRvD.102f2005A}
---. 2020{\natexlab{b}}, \prd, 102, 062005, \dodoi{10.1103/PhysRevD.102.062005}

\bibitem[{{Pierre Auger Collaboration}(2021)}]{PierreAuger:2021hun}
---. 2021, Eur. Phys. J. C, 81, 966, \dodoi{10.1140/epjc/s10052-021-09700-w}

\bibitem[{{Pierre Auger Collaboration}(2022)}]{PierreAuger:2022axr}
---. 2022, Astrophys. J., 935, 170, \dodoi{10.3847/1538-4357/ac7d4e}

\bibitem[{{Pierre Auger
  Collaboration}(2023{\natexlab{a}})}]{PierreAuger:2023fcr}
---. 2023{\natexlab{a}}, PoS, ICRC2023, 252, \dodoi{10.22323/1.444.0252}

\bibitem[{{Pierre Auger
  Collaboration}(2023{\natexlab{b}})}]{PierreAuger:2022atd}
---. 2023{\natexlab{b}}, JCAP, 05, 024, \dodoi{10.1088/1475-7516/2023/05/024}

\bibitem[{{Pierre Auger Collaboration}(2023{\natexlab{c}})}]{Berat:2023ttm}
---. 2023{\natexlab{c}}, EPJ Web Conf., 283, 06001,
  \dodoi{10.1051/epjconf/202328306001}

\bibitem[{{Pierre Auger
  Collaboration}(2023{\natexlab{d}})}]{PierreAuger:2023onx}
---. 2023{\natexlab{d}}, PoS, ICRC2023, 246, \dodoi{10.22323/1.444.0246}

\bibitem[{{Planck Collaboration} {et~al.}(2020){Planck Collaboration},
  {Aghanim}, {Akrami}, {Arroja}, {Ashdown}, {Aumont}, {Baccigalupi},
  {Ballardini}, {Banday}, {Barreiro}, {Bartolo}, {Basak}, {Battye}, {Benabed},
  {Bernard}, {Bersanelli}, {Bielewicz}, {Bock}, {Bond}, {Borrill}, {Bouchet},
  {Boulanger}, {Bucher}, {Burigana}, {Butler}, {Calabrese}, {Cardoso},
  {Carron}, {Casaponsa}, {Challinor}, {Chiang}, {Colombo}, {Combet},
  {Contreras}, {Crill}, {Cuttaia}, {de Bernardis}, {de Zotti}, {Delabrouille},
  {Delouis}, {D{\'e}sert}, {Di Valentino}, {Dickinson}, {Diego}, {Donzelli},
  {Dor{\'e}}, {Douspis}, {Ducout}, {Dupac}, {Efstathiou}, {Elsner},
  {En{\ss}lin}, {Eriksen}, {Falgarone}, {Fantaye}, {Fergusson},
  {Fernandez-Cobos}, {Finelli}, {Forastieri}, {Frailis}, {Franceschi},
  {Frolov}, {Galeotta}, {Galli}, {Ganga}, {G{\'e}nova-Santos}, {Gerbino},
  {Ghosh}, {Gonz{\'a}lez-Nuevo}, {G{\'o}rski}, {Gratton}, {Gruppuso},
  {Gudmundsson}, {Hamann}, {Handley}, {Hansen}, {Helou}, {Herranz},
  {Hildebrandt}, {Hivon}, {Huang}, {Jaffe}, {Jones}, {Karakci}, {Keih{\"a}nen},
  {Keskitalo}, {Kiiveri}, {Kim}, {Kisner}, {Knox}, {Krachmalnicoff}, {Kunz},
  {Kurki-Suonio}, {Lagache}, {Lamarre}, {Langer}, {Lasenby}, {Lattanzi},
  {Lawrence}, {Le Jeune}, {Leahy}, {Lesgourgues}, {Levrier}, {Lewis},
  {Liguori}, {Lilje}, {Lilley}, {Lindholm}, {L{\'o}pez-Caniego}, {Lubin}, {Ma},
  {Mac{\'\i}as-P{\'e}rez}, {Maggio}, {Maino}, {Mandolesi}, {Mangilli},
  {Marcos-Caballero}, {Maris}, {Martin}, {Martinelli},
  {Mart{\'\i}nez-Gonz{\'a}lez}, {Matarrese}, {Mauri}, {McEwen}, {Meerburg},
  {Meinhold}, {Melchiorri}, {Mennella}, {Migliaccio}, {Millea}, {Mitra},
  {Miville-Desch{\^e}nes}, {Molinari}, {Moneti}, {Montier}, {Morgante}, {Moss},
  {Mottet}, {M{\"u}nchmeyer}, {Natoli}, {N{\o}rgaard-Nielsen}, {Oxborrow},
  {Pagano}, {Paoletti}, {Partridge}, {Patanchon}, {Pearson}, {Peel}, {Peiris},
  {Perrotta}, {Pettorino}, {Piacentini}, {Polastri}, {Polenta}, {Puget},
  {Rachen}, {Reinecke}, {Remazeilles}, {Renault}, {Renzi}, {Rocha}, {Rosset},
  {Roudier}, {Rubi{\~n}o-Mart{\'\i}n}, {Ruiz-Granados}, {Salvati}, {Sandri},
  {Savelainen}, {Scott}, {Shellard}, {Shiraishi}, {Sirignano}, {Sirri},
  {Spencer}, {Sunyaev}, {Suur-Uski}, {Tauber}, {Tavagnacco}, {Tenti},
  {Terenzi}, {Toffolatti}, {Tomasi}, {Trombetti}, {Valiviita}, {Van Tent},
  {Vibert}, {Vielva}, {Villa}, {Vittorio}, {Wandelt}, {Wehus}, {White},
  {White}, {Zacchei}, \& {Zonca}}]{2020A&A...641A...1P}
{Planck Collaboration}, {Aghanim}, N., {Akrami}, Y., {et~al.} 2020, \aap, 641,
  A1, \dodoi{10.1051/0004-6361/201833880}

\bibitem[{Skrutskie {et~al.}(2006)}]{2MASS:2006qir}
Skrutskie, M.~F., {et~al.} 2006, Astron. J., 131, 1163, \dodoi{10.1086/498708}

\bibitem[{{Telescope Array Collaboration}(2013)}]{TelescopeArray:2012uws}
{Telescope Array Collaboration}. 2013, Nucl. Instrum. Meth. A, 689, 87,
  \dodoi{10.1016/j.nima.2012.05.079}

\bibitem[{{Telescope Array Collaboration}(2020)}]{TelescopeArray:2020cbq}
---. 2020, Astrophys. J. Lett., 898, L28, \dodoi{10.3847/2041-8213/aba0bc}

\bibitem[{Tinyakov \& Urban(2015)}]{Tinyakov:2014fwa}
Tinyakov, P.~G., \& Urban, F.~R. 2015, J. Exp. Theor. Phys., 120, 533,
  \dodoi{10.1134/S1063776115030231}

\bibitem[{Unger \& Farrar(2024)}]{Unger:2023lob}
Unger, M., \& Farrar, G.~R. 2024, Astrophys. J., 970, 95,
  \dodoi{10.3847/1538-4357/ad4a54}

\bibitem[{{Zatsepin} \& {Kuz'min}(1966)}]{1966JETPL...4...78Z}
{Zatsepin}, G.~T., \& {Kuz'min}, V.~A. 1966, J.\ Exp.\ Theor.\ Phys.\ Lett., 4,
  78

\end{thebibliography}
\bibliographystyle{aasjournal}

\newpage

\AuthorCollaborationLimit=3000
{\bf\Large{The Pierre Auger Collaboration}}
\\
A.~Abdul Halim$^{13}$,
P.~Abreu$^{70}$,
M.~Aglietta$^{53,51}$,
I.~Allekotte$^{1}$,
K.~Almeida Cheminant$^{78,77}$,
A.~Almela$^{7,12}$,
R.~Aloisio$^{44,45}$,
J.~Alvarez-Mu\~niz$^{76}$,
A.~Ambrosone$^{44}$,
J.~Ammerman Yebra$^{76}$
% ...

\renewcommand*\descriptionlabel[1]{\hspace\labelsep\normalfont #1}
\begin{description}[labelsep=0.2em,align=right,labelwidth=0.7em,labelindent=0em,leftmargin=2em,noitemsep,before={\renewcommand\makelabel[1]{##1 }}]
\item[$^{1}$] Centro At\'omico Bariloche and Instituto Balseiro (CNEA-UNCuyo-CONICET), San Carlos de Bariloche, Argentina
\item[$^{2}$] Departamento de F\'\i{}sica and Departamento de Ciencias de la Atm\'osfera y los Oc\'eanos, FCEyN, Universidad de Buenos Aires and CONICET, Buenos Aires, Argentina
\item[$^{3}$] IFLP, Universidad Nacional de La Plata and CONICET, La Plata, Argentina
\item[$^{4}$] Instituto de Astronom\'\i{}a y F\'\i{}sica del Espacio (IAFE, CONICET-UBA), Buenos Aires, Argentina
\item[$^{5}$] Instituto de F\'\i{}sica de Rosario (IFIR) -- CONICET/U.N.R.\ and Facultad de Ciencias Bioqu\'\i{}micas y Farmac\'euticas U.N.R., Rosario, Argentina
\item[$^{6}$] Instituto de Tecnolog\'\i{}as en Detecci\'on y Astropart\'\i{}culas (CNEA, CONICET, UNSAM), and Universidad Tecnol\'ogica Nacional -- Facultad Regional Mendoza (CONICET/CNEA), Mendoza, Argentina
\item[$^{7}$] Instituto de Tecnolog\'\i{}as en Detecci\'on y Astropart\'\i{}culas (CNEA, CONICET, UNSAM), Buenos Aires, Argentina
\item[$^{8}$] International Center of Advanced Studies and Instituto de Ciencias F\'\i{}sicas, ECyT-UNSAM and CONICET, Campus Miguelete -- San Mart\'\i{}n, Buenos Aires, Argentina
\item[$^{9}$] Laboratorio Atm\'osfera -- Departamento de Investigaciones en L\'aseres y sus Aplicaciones -- UNIDEF (CITEDEF-CONICET), Argentina
\item[$^{10}$] Observatorio Pierre Auger, Malarg\"ue, Argentina
\item[$^{11}$] Observatorio Pierre Auger and Comisi\'on Nacional de Energ\'\i{}a At\'omica, Malarg\"ue, Argentina
\item[$^{12}$] Universidad Tecnol\'ogica Nacional -- Facultad Regional Buenos Aires, Buenos Aires, Argentina
\item[$^{13}$] University of Adelaide, Adelaide, S.A., Australia
\item[$^{14}$] Universit\'e Libre de Bruxelles (ULB), Brussels, Belgium
\item[$^{15}$] Vrije Universiteit Brussels, Brussels, Belgium
\item[$^{16}$] Centro Brasileiro de Pesquisas Fisicas, Rio de Janeiro, RJ, Brazil
\item[$^{17}$] Centro Federal de Educa\c{c}\~ao Tecnol\'ogica Celso Suckow da Fonseca, Petropolis, Brazil
\item[$^{18}$] Instituto Federal de Educa\c{c}\~ao, Ci\^encia e Tecnologia do Rio de Janeiro (IFRJ), Brazil
\item[$^{19}$] Universidade de S\~ao Paulo, Escola de Engenharia de Lorena, Lorena, SP, Brazil
\item[$^{20}$] Universidade de S\~ao Paulo, Instituto de F\'\i{}sica de S\~ao Carlos, S\~ao Carlos, SP, Brazil
\item[$^{21}$] Universidade de S\~ao Paulo, Instituto de F\'\i{}sica, S\~ao Paulo, SP, Brazil
\item[$^{22}$] Universidade Estadual de Campinas (UNICAMP), IFGW, Campinas, SP, Brazil
\item[$^{23}$] Universidade Estadual de Feira de Santana, Feira de Santana, Brazil
\item[$^{24}$] Universidade Federal de Campina Grande, Centro de Ciencias e Tecnologia, Campina Grande, Brazil
\item[$^{25}$] Universidade Federal do ABC, Santo Andr\'e, SP, Brazil
\item[$^{26}$] Universidade Federal do Paran\'a, Setor Palotina, Palotina, Brazil
\item[$^{27}$] Universidade Federal do Rio de Janeiro, Instituto de F\'\i{}sica, Rio de Janeiro, RJ, Brazil
\item[$^{28}$] Universidad de Medell\'\i{}n, Medell\'\i{}n, Colombia
\item[$^{29}$] Universidad Industrial de Santander, Bucaramanga, Colombia
\item[$^{30}$] Charles University, Faculty of Mathematics and Physics, Institute of Particle and Nuclear Physics, Prague, Czech Republic
\item[$^{31}$] Institute of Physics of the Czech Academy of Sciences, Prague, Czech Republic
\item[$^{32}$] Palacky University, Olomouc, Czech Republic
\item[$^{33}$] CNRS/IN2P3, IJCLab, Universit\'e Paris-Saclay, Orsay, France
\item[$^{34}$] Laboratoire de Physique Nucl\'eaire et de Hautes Energies (LPNHE), Sorbonne Universit\'e, Universit\'e de Paris, CNRS-IN2P3, Paris, France
\item[$^{35}$] Univ.\ Grenoble Alpes, CNRS, Grenoble Institute of Engineering Univ.\ Grenoble Alpes, LPSC-IN2P3, 38000 Grenoble, France
\item[$^{36}$] Universit\'e Paris-Saclay, CNRS/IN2P3, IJCLab, Orsay, France
\item[$^{37}$] Bergische Universit\"at Wuppertal, Department of Physics, Wuppertal, Germany
\item[$^{38}$] Karlsruhe Institute of Technology (KIT), Institute for Experimental Particle Physics, Karlsruhe, Germany
\item[$^{39}$] Karlsruhe Institute of Technology (KIT), Institut f\"ur Prozessdatenverarbeitung und Elektronik, Karlsruhe, Germany
\item[$^{40}$] Karlsruhe Institute of Technology (KIT), Institute for Astroparticle Physics, Karlsruhe, Germany
\item[$^{41}$] RWTH Aachen University, III.\ Physikalisches Institut A, Aachen, Germany
\item[$^{42}$] Universit\"at Hamburg, II.\ Institut f\"ur Theoretische Physik, Hamburg, Germany
\item[$^{43}$] Universit\"at Siegen, Department Physik -- Experimentelle Teilchenphysik, Siegen, Germany
\item[$^{44}$] Gran Sasso Science Institute, L'Aquila, Italy
\item[$^{45}$] INFN Laboratori Nazionali del Gran Sasso, Assergi (L'Aquila), Italy
\item[$^{46}$] INFN, Sezione di Catania, Catania, Italy
\item[$^{47}$] INFN, Sezione di Lecce, Lecce, Italy
\item[$^{48}$] INFN, Sezione di Milano, Milano, Italy
\item[$^{49}$] INFN, Sezione di Napoli, Napoli, Italy
\item[$^{50}$] INFN, Sezione di Roma ``Tor Vergata'', Roma, Italy
\item[$^{51}$] INFN, Sezione di Torino, Torino, Italy
\item[$^{52}$] Istituto di Astrofisica Spaziale e Fisica Cosmica di Palermo (INAF), Palermo, Italy
\item[$^{53}$] Osservatorio Astrofisico di Torino (INAF), Torino, Italy
\item[$^{54}$] Politecnico di Milano, Dipartimento di Scienze e Tecnologie Aerospaziali , Milano, Italy
\item[$^{55}$] Universit\`a del Salento, Dipartimento di Matematica e Fisica ``E.\ De Giorgi'', Lecce, Italy
\item[$^{56}$] Universit\`a dell'Aquila, Dipartimento di Scienze Fisiche e Chimiche, L'Aquila, Italy
\item[$^{57}$] Universit\`a di Catania, Dipartimento di Fisica e Astronomia ``Ettore Majorana``, Catania, Italy
\item[$^{58}$] Universit\`a di Milano, Dipartimento di Fisica, Milano, Italy
\item[$^{59}$] Universit\`a di Napoli ``Federico II'', Dipartimento di Fisica ``Ettore Pancini'', Napoli, Italy
\item[$^{60}$] Universit\`a di Palermo, Dipartimento di Fisica e Chimica ''E.\ Segr\`e'', Palermo, Italy
\item[$^{61}$] Universit\`a di Roma ``Tor Vergata'', Dipartimento di Fisica, Roma, Italy
\item[$^{62}$] Universit\`a Torino, Dipartimento di Fisica, Torino, Italy
\item[$^{63}$] Benem\'erita Universidad Aut\'onoma de Puebla, Puebla, M\'exico
\item[$^{64}$] Unidad Profesional Interdisciplinaria en Ingenier\'\i{}a y Tecnolog\'\i{}as Avanzadas del Instituto Polit\'ecnico Nacional (UPIITA-IPN), M\'exico, D.F., M\'exico
\item[$^{65}$] Universidad Aut\'onoma de Chiapas, Tuxtla Guti\'errez, Chiapas, M\'exico
\item[$^{66}$] Universidad Michoacana de San Nicol\'as de Hidalgo, Morelia, Michoac\'an, M\'exico
\item[$^{67}$] Universidad Nacional Aut\'onoma de M\'exico, M\'exico, D.F., M\'exico
\item[$^{68}$] Institute of Nuclear Physics PAN, Krakow, Poland
\item[$^{69}$] University of \L{}\'od\'z, Faculty of High-Energy Astrophysics,\L{}\'od\'z, Poland
\item[$^{70}$] Laborat\'orio de Instrumenta\c{c}\~ao e F\'\i{}sica Experimental de Part\'\i{}culas -- LIP and Instituto Superior T\'ecnico -- IST, Universidade de Lisboa -- UL, Lisboa, Portugal
\item[$^{71}$] ``Horia Hulubei'' National Institute for Physics and Nuclear Engineering, Bucharest-Magurele, Romania
\item[$^{72}$] Institute of Space Science, Bucharest-Magurele, Romania
\item[$^{73}$] Center for Astrophysics and Cosmology (CAC), University of Nova Gorica, Nova Gorica, Slovenia
\item[$^{74}$] Experimental Particle Physics Department, J.\ Stefan Institute, Ljubljana, Slovenia
\item[$^{75}$] Universidad de Granada and C.A.F.P.E., Granada, Spain
\item[$^{76}$] Instituto Galego de F\'\i{}sica de Altas Enerx\'\i{}as (IGFAE), Universidade de Santiago de Compostela, Santiago de Compostela, Spain
\item[$^{77}$] IMAPP, Radboud University Nijmegen, Nijmegen, The Netherlands
\item[$^{78}$] Nationaal Instituut voor Kernfysica en Hoge Energie Fysica (NIKHEF), Science Park, Amsterdam, The Netherlands
\item[$^{79}$] Stichting Astronomisch Onderzoek in Nederland (ASTRON), Dwingeloo, The Netherlands
\item[$^{80}$] Universiteit van Amsterdam, Faculty of Science, Amsterdam, The Netherlands
\item[$^{81}$] Case Western Reserve University, Cleveland, OH, USA
\item[$^{82}$] Colorado School of Mines, Golden, CO, USA
\item[$^{83}$] Department of Physics and Astronomy, Lehman College, City University of New York, Bronx, NY, USA
\item[$^{84}$] Michigan Technological University, Houghton, MI, USA
\item[$^{85}$] New York University, New York, NY, USA
\item[$^{86}$] University of Chicago, Enrico Fermi Institute, Chicago, IL, USA
\item[$^{87}$] University of Delaware, Department of Physics and Astronomy, Bartol Research Institute, Newark, DE, USA
\item[] -----
\item[$^{a}$] Max-Planck-Institut f\"ur Radioastronomie, Bonn, Germany
\item[$^{b}$] also at Kapteyn Institute, University of Groningen, Groningen, The Netherlands
\item[$^{c}$] School of Physics and Astronomy, University of Leeds, Leeds, United Kingdom
\item[$^{d}$] Fermi National Accelerator Laboratory, Fermilab, Batavia, IL, USA
\item[$^{e}$] Pennsylvania State University, University Park, PA, USA
\item[$^{f}$] Colorado State University, Fort Collins, CO, USA
\item[$^{g}$] Louisiana State University, Baton Rouge, LA, USA
\item[$^{h}$] now at Graduate School of Science, Osaka Metropolitan University, Osaka, Japan
\item[$^{i}$] Institut universitaire de France (IUF), France
\end{description}

\end{document}